\documentclass[3p,times,10pt]{elsarticle}
\usepackage{amsfonts}
\usepackage{amssymb}
\usepackage{amsmath}
\usepackage{graphicx}
\usepackage{epsfig}
\usepackage{ctable}

\begin{document}

\begin{frontmatter}

\title{Covariant gaussian approximation in Ginzburg - Landau model}

\author[PKU,CICQ]{J.F. Wang}
\author[PKU,CICQ]{D.P. Li}
\author[NTNU]{H.C. Kao}
\author[NCTU]{B. Rosenstein}

\address[PKU]{School of Physics, Peking University, Beijing 100871, China}
\address[CICQ]{Collaborative Innovation Center of Quantum Matter, Beijing 100871, China}
\address[NTNU]{Physics Department, National Taiwan Normal University, Taipei 11677, Taiwan, R.O.C.}
\address[NCTU]{Electrophysics Department, National Chiao Tung University, Hsinchu 30050, Taiwan, R.O.C.}

\date{\today}

\begin{abstract}
Condensed matter systems undergoing second order transition away from the
critical fluctuation region are usually described sufficiently well by the
mean field approximation. The critical fluctuation region, determined by
the Ginzburg criterion, $\left \vert T/T_{c}-1\right \vert \ll Gi$, is narrow
even in high $T_{c}$ superconductors and has universal features well
captured by the renormalization group method. However recent experiments on
magnetization, conductivity and Nernst effect suggest that fluctuations
effects are large in a wider region both above and below $T_{c}$. In
particular some ``pseudogap" phenomena and strong renormalization of the mean
field critical temperature $T_{mf}$ can be interpreted as strong fluctuations
effects that are nonperturbative (cannot be accounted for by ``gaussian
fluctuations"). The physics in a broader region therefore requires more
accurate approach. Self consistent methods are generally ``non - conserving" in
the sense that the Ward identities are not obeyed. This is especially
detrimental in the symmetry broken phase where, for example, Goldstone
bosons become massive. Covariant gaussian approximation remedies these
problems. The Green's functions obey all the Ward identities and describe
the fluctuations much better. The results for the order parameter correlator
and magnetic penetration depth of the Ginzburg - Landau model of
superconductivity are compared with both Monte Carlo simulations and experiments
in high $T_{c}$ cuprates.
\end{abstract}

\begin{keyword}
Covariant gaussian approximation \sep Ginzburg - Landau \sep Ward identity \sep superconducting thermal fluctuations \sep magnetic penetration depth
\end{keyword}

\end{frontmatter}

\section{Introduction}

Many - body systems at nonzero temperature exhibit a host of physical
phenomena triggered by strong thermal fluctuations\cite{Lubenskybook}. Most
remarkable are of course phase transitions, in which the ground state
rearranges. In addition, crossovers and qualitatively recognizable
phenomena, such as metastability or possibly the ``pseudogap" physics in high
$T_{c}$ cuprates, also attract intensive interest. It was noticed by Landau
that for temperatures close to critical point of a second order transition,
symmetry and energetics considerations suggest that the system can be
described by a rather simple universal Ginzburg - Landau (GL) model. The
model expressed in terms of an appropriate order parameter field $\phi $
contains a very small number of phenomenological parameters, since one
retains only terms up to fourth power of $\phi $ in the GL\ free energy (on
the condition that it is well separated from the tricritical point where the
sixth order of $\phi $ cannot be neglected).

The universality was established in the critical fluctuation region, where
physics is determined by several critical exponents well captured by the
renormalization group approach to the GL $\phi ^{4}$ theory\cite{Amit}.
However the width of the critical region, determined by the Ginzburg
criterion\cite{Varlamov}, $\left\vert T/T_{c}-1\right\vert \ll Gi$, is very
narrow even in the most fluctuating materials like the high $T_{c}$
superconductors for which the Ginzburg number $Gi$ is of order $0.05$. Away
from the critical fluctuation region, the universality might not hold. One
thus traditionally resorts to a variety of mean field or self - consistent
models. These models focus on different degrees of freedom, and so a
statistical system is often represented in several different ways depending
on the choice of the quantity that is considered self - consistently.
Examples of self - consistent approaches range from the Bragg - Williams
approximation for spin systems to the BCS approximation of conventional
superconductors and the Hartree - Fock approximation in the electron gas or
liquid. Thus the mean field approach is essentially different from the
economic GL description in terms of the order parameter field $\phi $
defined solely by the symmetry properties of the system. A question arises
whether the GL model can nevertheless be reliable away from its intended
range of applicability near the criticality.

The GL approach in terms of the order parameter however is in fact widely
and successfully used for description of the fluctuations outside the
critical region\cite{RMP}. There are several arguments why actually one can
use the universal model beyond its original range of applicability\cite%
{Lubenskybook,Varlamov,1999}. Very often the fluctuations are accounted for
by the gaussian fluctuations approximation to the GL model\cite{Tinkham}.
This completely neglects the $\phi ^{4}$ coupling of the order parameter
that sometimes is included perturbatively\cite{Varlamov}. Various
approximations beyond perturbation theory were developed.

Recent experiments on magnetization\cite{magnetization}, conductivity\cite%
{conductivity} and Nernst effect\cite{Ong} of high $T_{c}$ superconductors
suggest that fluctuation effects are significant and even large in a much
wider region both above and below $T_{c}$. First of all, the renormalization
of the critical temperature $T_{c}$ from its ``microscopic" or ``mean field"
value, $T_{mf}$, is non - universal and might be of the same order of
magnitude as $T_{c}$. Superconducting fluctuations as far above $T_{c}$ as
$2T_{c}$ still dominate over the normal state contributions to the above
three physical phenomena. Similarly below $T_{c}$ (sometimes even close to
zero temperature and definitely far away from the critical fluctuation
region) the fluctuations determine the nontrivial penetration depth
dependence on temperature\cite{Shih-Fu Lee,Broun,Yuri} and magnetic
properties\cite{RMP}. In particular some of the still poorly understood
phenomena in underdoped and optimally doped cuprates known as the
``pseudogap" features\cite{pseudogap} can be interpreted as strong
fluctuations effects that are nonperturbative\cite{Jiang} (cannot be
accounted for by gaussian fluctuations) rather than stemming from a yet
unknown microscopic origin.

It has been noted over the years that despite the anticipated breakdown of
the universality far beyond the critical fluctuation region strong
fluctuations are still well captured by the interacting $\phi ^{4}$ GL theory%
\cite{Varlamov,RMP}. Of course some approximations have to be made and the
most natural is a self - consistent method for the expectation value and the
correlator of the order parameter field known as gaussian approximation\cite%
{Stevenson}. As in most of the standard variants of the self - consistent
mean field theories, within the gaussian approximation one encounters a
difficulty with preserving the basic symmetries of the problem. In the
symmetric phase gaussian approximation (mean field in terms of the order
parameter and its correlator) is generally symmetry conserving and thus
sufficient for qualitative and even quantitative description. However below
the transition temperature this is not the case. The symmetry is not
preserved within the ``naive" gaussian approximation and as a result even
such a basic phenomenon as appearance of massless Goldstone modes in the
symmetry broken phase cannot be accounted for. The excitation becomes
massive and thus a symmetry conserving approach is required\cite{Baym}.

In this paper we show that a special uniquely defined variant of the mean
field approximation, the covariant gaussian approximation (CGA) around the
spontaneous symmetry broken state satisfies this requirement. The
approximation is conserving in that it preserves the symmetry properties
of the correlators of the order parameter characterizing the ordered phase.
The CGA Green's functions obey all the Ward identities and describe the
fluctuations much better. The results for the two and four point correlators
of the Ginzburg - Landau model are compared with Monte Carlo simulations and
experiments in superconductors.

The paper is organized as follows. In section 2 the covariant gaussian
approximation in the symmetry broken phase is developed using the simplest
toy model: the one dimensional spin chain equivalent mathematically to the
double well anharmonic oscillator in quantum mechanics. The Dyson -
Schwinger (DS) equations method is used and results for the two and four
point correlators are compared with direct numerical calculations, Monte
Carlo (MC) simulations as well as perturbation theory. In section 3 we apply
CGA to the $O\left( 2\right) $ invariant model and show how the
corresponding Ward identities are derived. In section 4 the cluster Monte
Carlo method appropriate for the calculation inside the symmetry broken
phase is briefly described. The correlator of the order parameter of
isotropic GL model in 3D is calculated with CGA. In section 5 we derive the
expression for the magnetic penetration depth of an anisotropic
superconductor and compare the CGA results with direct MC simulations and
experimental data on penetration depth of several high $T_{c}$ cuprates and
discuss the results. Conclusions are given in section 6.

\section{Hierarchy of variational conserving approximations defined as
truncations of DS equations}

\subsection{A toy model and basic definitions}

To present the approximation scheme we will make use of the simplest
nontrivial model: statistical physics of a one dimensional classical chain
that is equivalent to the quantum mechanics of the anharmonic oscillator.
Our starting point therefore will be the following free energy as a function
of a single component order parameter (equivalently Euclidean (Matsubara)
action of an anharmonic oscillator):%
\begin{equation}
A=\frac{1}{\omega }\int dx\left\{ \frac{{1}}{2}\left( \partial \phi \left(
x\right) \right) ^{2}+\frac{a}{2}\phi \left( x\right) ^{2}+\frac{1}{4}\phi
\left( x\right) ^{4}-J\left( x\right) \phi \left( x\right) \right\} \text{.}
\label{actionZ2}
\end{equation}%
The dimensionless coefficient $\omega $, proportional to temperature $T$ in
statistical physics of classical chain or to $\hbar $ in quantum anharmonic
oscillator, determines the strength of the (thermal or quantum)
fluctuations. The normalization of the order parameter field $\phi \left(
x\right) $ and the position $x$ is such that $\omega $ is dimensionless
(proportional to $\sqrt{Gi}$ where $Gi$ is the Ginzburg number\cite{RMP})
and the coefficient of interaction term is $1/4$. This free energy without
external source ($J\left( x\right) =0$) has $Z_{2}$ symmetry, that is,
invariant under $\phi \left( x\right) \rightarrow -\phi \left( x\right) $.
The thermodynamics of this model is determined by the statistical sum
\begin{equation}
Z=\int \mathcal{D}\phi \left( x\right) e^{-A\left[ \phi \left( x\right) %
\right] }\text{.}  \label{statsum}
\end{equation}%
The $n$-body correlator, the main object of interest in the present paper,
is defined by%
\begin{equation}
\left\langle \phi \left( x_{1}\right) \cdots \phi \left( x_{n}\right)
\right\rangle =\left. Z^{-1}\int_{\phi \left( x\right) }\phi \left(
x_{1}\right) \cdots \phi \left( x_{n}\right) e^{-A\left[ \phi \left(
x\right) \right] }\right\vert _{J=0}\text{.}  \label{corrdef}
\end{equation}%
This full $n$-body correlator can be conveniently written\cite{Amit} in
terms of connected correlators denoted by $\left\langle \cdots \right\rangle
_{c}$, while the later can be conveniently expressed via ``cumulants" $\Gamma
\left( x_{1},\cdots ,x_{n}\right) $. For example the simplest two body
correlator can be written as%
\begin{equation}
\left\langle \phi \left( x\right) \phi \left( y\right) \right\rangle
=\left\langle \phi \left( x\right) \right\rangle \left\langle \phi \left(
y\right) \right\rangle +\left\langle \phi \left( x\right) \phi \left(
y\right) \right\rangle _{c}\equiv \varphi \left( x\right) \varphi \left(
y\right) +G\left( x,y\right) =\varphi \left( x\right) \varphi \left(
y\right) +\Gamma ^{-1}\left( x,y\right)  \label{twobodyCorr}
\end{equation}%
Here the order parameter average is denoted by the ``classical field" $%
\varphi \left( x\right) \equiv \left\langle \phi \left( x\right)
\right\rangle $ and the two - body cumulant $\Gamma \left( x,y\right) $ is
the inverse in the matrix sense to the connected correlator%
\begin{equation}
\int_{u}\Gamma \left( x,u\right) G\left( u,y\right) =\delta \left(
x-y\right) \text{.}  \label{inversedef}
\end{equation}%
Similarly the\ three - body connected correlator is expressed via cumulants
as%
\begin{equation}
\left\langle \phi \left( x\right) \phi \left( y\right) \phi \left( z\right)
\right\rangle _{c}\equiv G_{3}\left( x,y,z\right) =-\int_{u,v,w}G\left(
x,u\right) G\left( y,v\right) G\left( z,w\right) \Gamma \left( u,v,w\right)
\text{.}  \label{connectedcorr}
\end{equation}

From the functional integral in the presence of source $J\left( x\right) $,
one can derive the equation of state\ (ES), the first of the infinite set of
coupled DS equations for connected correlators :%
\begin{equation}
J\left( x\right) =\left( -\partial ^{2}+a\right) \varphi \left( x\right)
+\varphi \left( x\right) ^{3}+3\varphi \left( x\right) G\left( x,x\right)
+G_{3}\left( x,x,x\right) \text{.}  \label{exactES}
\end{equation}%
To simplify the expression, we will set $\omega =1$ from now on. This always
can be achieved by rescaling in 1D system since there are no ultraviolet
(UV) or infrared (IR) divergences. The parameter $\omega $ will be
reinstated when we need to use it to count the order of ``loop expansion".

The second DS equation is the functional derivative of the equation of state,%
\begin{eqnarray}
\frac{\delta J\left( x\right) }{\delta \phi \left( y\right) } &\equiv
&\Gamma \left( x,y\right) =\left( -\partial _{x}^{2}+a\right) \delta \left(
x-y\right) +3\delta \left( x-y\right) \varphi \left( x\right) ^{2}+3\delta
\left( x-y\right) G\left( x,x\right)  \label{DS2exact} \\
&&+3\varphi \left( x\right) \int_{u}\Gamma \left( y,u\right) G_{3}\left(
x,x,u\right) +\int_{u}\Gamma \left( y,u\right) G_{4}\left( x,x,x,u\right)
\text{.}  \notag
\end{eqnarray}%
The infinite set of equations is not useful unless a way to decouple higher
order equations is proposed. The simplest one is the classical approximation.

\subsection{Classical approximation and its variational interpretation}

The classical approximation consists of neglecting the two and three body
correlators in the equation of state, Eq.(\ref{exactES}),
\begin{equation}
J\left( x\right) =\left( -\partial ^{2}+a\right) \varphi \left( x\right)
+\varphi \left( x\right) ^{3}\text{,}  \label{offcl}
\end{equation}%
so that the second and higher equations are decoupled from the first. Then
the ``minimization equation", that is just the on-shell ($J=0$) ES,%
\begin{equation}
\left( -\partial ^{2}+a\right) \varphi \left( x\right) +\varphi \left(
x\right) ^{3}=0,  \label{classical}
\end{equation}%
is solved.

For $a<0$ there are typically several solutions of this equation.
Restricting ourselves to the translational invariant ones (namely excluding
kinks), one has
\begin{equation}
\varphi \left( x\right) =0\text{,}\pm v;\text{ \ \ }v=\sqrt{-a}\text{.}
\label{classicalsol}
\end{equation}%
Despite the fact that we know there is no spontaneous symmetry breaking in $%
d=1$, a rather precise approximation scheme for a $Z_{2}$ invariant quantity
emerges when one of the two ``would be" symmetry broken solutions, say $%
\varphi \left( x\right) =+v$, is considered in the intermediate steps.

Note that despite the fact that the minimization principle involved only the
one - body cumulant, $\varphi \left( x\right) $, one can still calculate the
higher cumulants within the classical approximation. These are given by
functional derivatives of the source $J\left( x\right) $ with respect to $%
\varphi \left( x\right) $ in truncated ES Eq.(\ref{offcl}):
\begin{subequations}
\begin{eqnarray}
\Gamma _{cl}\left( x,y\right) &=&\frac{\delta J\left( x\right) }{\delta
\varphi \left( y\right) }=(-\partial _{x}^{2}+a+3\varphi \left( x\right)
^{2})\delta \left( x-y\right) ;  \label{1} \\
\Gamma _{cl}^{3}\left( x,y,z\right) &=&\frac{\delta ^{2}J\left( x\right) }{%
\delta \varphi \left( y\right) \delta \varphi \left( z\right) }=6\delta
\left( x-y\right) \delta \left( x-z\right) \varphi \left( x\right) ;
\label{2} \\
\Gamma _{cl}^{4}\left( x,y,z,u\right) &=&\frac{\delta ^{3}J\left( x\right) }{%
\delta \varphi \left( y\right) \delta \varphi \left( z\right) \delta \varphi
\left( u\right) }=6\delta \left( x-y\right) \delta \left( x-z\right) \delta
\left( x-u\right) \text{.}  \label{3}
\end{eqnarray}%
The cumulants beyond the fourth order vanish within this approximation.

The full correlator, a quantity invariant under the $Z_{2}$, in this
approximation is:
\end{subequations}
\begin{equation}
P\left( x\right) \equiv \left\langle \phi \left( x\right) \phi \left(
0\right) \right\rangle =v^{2}+\Gamma _{cl}^{-1}\left( x,0\right) \text{.}
\label{Invcorrdef}
\end{equation}%
The matrix inversion of
\begin{equation}
\Gamma _{cl}\left( x,y\right) =\left( -\partial _{x}^{2}+a+3v^{2}\right)
\delta \left( x-y\right) =\left( -\partial _{x}^{2}+m^{2}\right) \delta
\left( x-y\right)  \label{classprop}
\end{equation}%
as usual is performed in Fourier space, $\phi
_{k}=\sum\nolimits_{k}e^{-ikx}\phi \left( x\right) $. Here the mass is $%
m^{2}=2v^{2}$. Therefore the full correlator in momentum space is just%
\begin{equation}
P\left( k\right) =\left\langle \phi _{k}\phi _{-k}\right\rangle =v^{2}\delta
_{k}+\frac{1}{k^{2}+m^{2}}\text{.}  \label{classGk}
\end{equation}

It can be corrected to ``one loop" by calculating the ``tadpole" diagram and
takes the form%
\begin{equation}
P\left( k\right) =v^{2}\delta _{k}+\omega \left\{ 2v\left\langle \phi
_{k}\right\rangle +G_{cl}\left( k\right) \right\} =v^{2}\delta _{k}+\omega
\left\{ -\delta _{k}\sum\nolimits_{p}\frac{1}{p^{2}+m^{2}}+\frac{1}{%
k^{2}+m^{2}}\right\} \text{,}
\end{equation}%
where the fluctuation parameter $\omega $ has been reinstated. Returning to
the configuration space, one has
\begin{equation}
P\left( x\right) =v^{2}+\frac{\omega }{2\pi }\int_{k=-\infty }^{\infty }%
\frac{e^{ikx}-1}{k^{2}+m^{2}}=v^{2}+\frac{\omega }{2m}\left( e^{-m\left\vert
x\right\vert }-1\right) \text{.}  \label{classGx}
\end{equation}%
This is compared in Figure 1 (shown as dashed blue lines) with the
numerically calculated correlator (see Appendix A) and cluster Monte Carlo
simulations (see details in section 4.2) for negative $a$ and $\omega =1$.
One notices that the classical approximation exhibits long range order at
negative $a$, which is definitely wrong in the case of one dimension as
clearly shown by both numerical calculation and MC simulations. The
classical approximation that was obtained by an ad hoc truncation of the
exact ES can be made a part of a systematically improvable scheme by
considering the arbitrarily omitted last two terms in Eq.(\ref{exactES})
as a perturbation.

\begin{figure}[t]
\centering
\includegraphics[width=8cm]{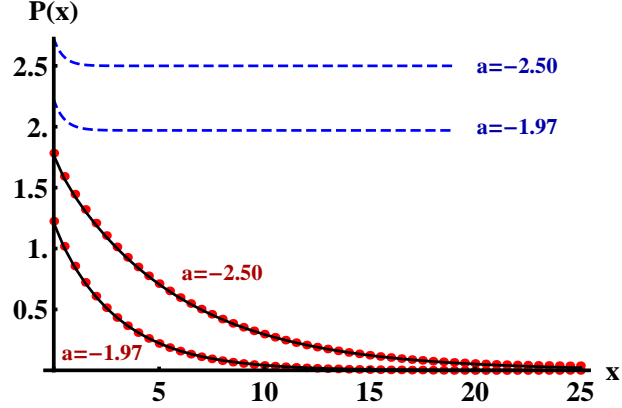}
\caption{Correlator in real space $P\left( x\right) $ calculated by
different methods. The parameter $\protect\omega $ is taken to be 1 (can be
scaled away) and values of $a=-1.97,-2.5$ are given. Black lines are direct
spectrum calculations (see Appendix A). Red dots are Monte Carlo simulations
with error bars smaller than the size of dots. Dashed blue lines are the
classical approximation, Eq.(\protect\ref{classGx}).}
\label{fig1}
\end{figure}

\begin{figure}[t!]
\centering
\includegraphics[width=16cm]{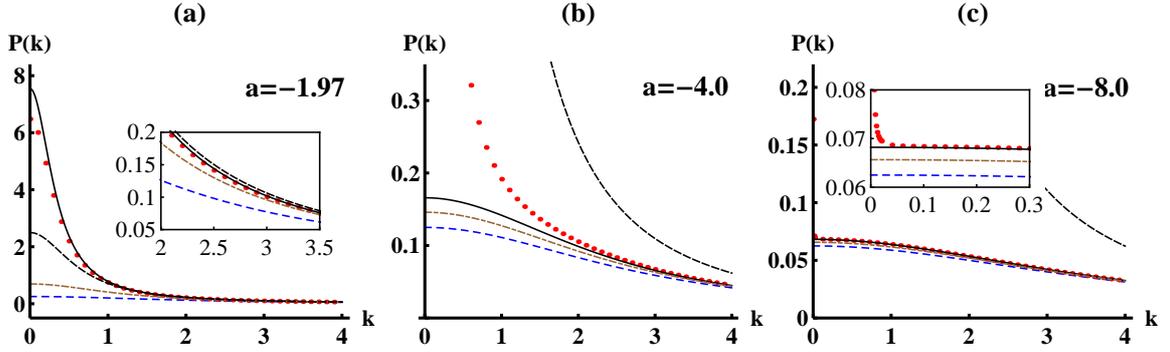}
\caption{Correlator in momentum space calculated by different methods at $%
a=-1.97,-4,-8$. Red dots are direct spectrum calculations (see Appendix A).
Dashed blue line is the classical approximation, Eq.(\protect\ref{classGk}).
The black line is CGA around broken phase solution of the minimization
equations, Eq.(\protect\ref{opcorrelator}). Dashed black is gaussian
approximation in the symmetric phase, Eq.(\protect\ref{gtr}). Brown dashed
dot line is the naive gaussian approximation in the ``would be" broken phase,
Eq.(\protect\ref{corrasym}). The erroneous delta function part at zero
momentum is not shown in this figure. Insets in (a) and (c) show
amplifications of the corresponding small regions.}
\label{fig2}
\end{figure}

The minimization equation Eq.(\ref{classical}) can be interpreted
variationally as minimizing the quantum mechanical double well Hamiltonian
\begin{equation}
H=-\frac{1}{2}\frac{\partial ^{2}}{\partial \phi ^{2}}+\frac{a}{2}\phi ^{2}+%
\frac{1}{4}\phi ^{4}\text{,}  \label{Hamiltonian}
\end{equation}%
on the set of coherent states wave functions (become functionals in higher
dimensional models):%
\begin{equation}
\psi \left( \phi \right) =\left( \frac{1}{2\pi g}\right) ^{1/4}\exp \left[ -%
\frac{1}{4g}\left( \phi -\varphi \right) ^{2}\right] \text{.}
\label{wavefunction}
\end{equation}%
The parameter $g$, square of the width of the gaussian wave function, is
arbitrary but fixed. The expectation value of energy%
\begin{equation}
E\left( \varphi ,g\right) =\left\langle \psi \left( \phi \right) \left\vert
H\right\vert \psi \left( \phi \right) \right\rangle =\frac{1}{8g}+\frac{a}{2}%
g+\frac{3}{4}g^{2}+\frac{1}{2}\left( a+3g\right) \varphi ^{2}+\frac{1}{4}%
\varphi ^{4}  \label{Energy}
\end{equation}%
is minimized with respect to the shift $\varphi $. This leads to the
translational invariant form of the classical equation Eq.(\ref{classical})
for small $g$ (localized gaussian).

In principle one can do better. Variationally one can optimize not just the
shift of the gaussian wave function, but also the width $g$.

\subsection{Gaussian variational principle}

Optimization of the energy Eq.(\ref{Energy}) with respect to both parameters
$\varphi $\ and $g$ would give us the minimization equations of the
following form\cite{Stevenson}%
\begin{equation}
\frac{\delta }{\delta \varphi }E\left( \varphi ,g\right) =\left( a+3g\right)
\varphi +\varphi ^{3}=0;  \label{minimshift}
\end{equation}%
\begin{equation}
\frac{\delta }{\delta g}E\left( \varphi ,g\right) =-\frac{1}{8g^{2}}+\frac{a%
}{2}+\frac{3}{2}g+\frac{3}{2}\varphi ^{2}=0\text{.}  \label{minimgap}
\end{equation}%
We will see these two equations coincide with the ``two - body" truncation of
the DS equations for $J\left( x\right) =0$ under the assumption of
translational invariance.

Instead of truncating out the two-body and higher cumulant, like in the
classical approximation, one can truncate the DS set of equations starting
from the three-body cumulant. Of course the approximation becomes more
complicated. Indeed let us truncate the two lowest order DS equations, Eq.(%
\ref{exactES}) and Eq.(\ref{DS2exact}), leaving out only the last term in
ES, Eq.(\ref{exactES})%
\begin{equation}
J\left( x\right) =\left( -\partial ^{2}+a\right) \varphi \left( x\right)
+\varphi \left( x\right) ^{3}+3\varphi \left( x\right) G^{tr}\left(
x,x\right) \text{,}  \label{shifteq}
\end{equation}%
and the last two terms of the second DS equation, Eq.(\ref{DS2exact}),%
\begin{equation}
\Gamma ^{tr}\left( x,y\right) =\left( -\partial _{x}^{2}+a\right) \delta
\left( x-y\right) +3\delta \left( x-y\right) \varphi \left( x\right)
^{2}+3\delta \left( x-y\right) G^{tr}\left( x,x\right) \text{.}
\label{gapeq}
\end{equation}%
Here the superscript ``tr" indicates that the correlators satisfy the
approximated truncated equations instead of the exact ones.

In the naive gaussian variational principle approach\cite{Stevenson}, one
considers these equations as minimization equations with $\varphi \left(
x\right) $ and $G^{tr}\left( x,y\right) $ identified as the connected
correlator inverse to $\Gamma ^{tr}\left( x,y\right) $ in the matrix sense.
Let us assume the space translational invariance leading to%
\begin{equation}
\varphi \left( x\right) =\varphi ,\text{\ }G^{tr}\left( x,y\right)
=g^{tr}\left( x-y\right) =\sum\nolimits_{k}e^{ik\left( x-y\right) }g_{k}^{tr}%
\text{,}  \label{translationsym}
\end{equation}%
so that the minimization equations for $J=0$ take the form%
\begin{equation}
0=a\varphi +\varphi ^{3}+3\varphi \sum\nolimits_{k}g_{k}^{tr}\text{,}
\label{shift1}
\end{equation}%
and%
\begin{equation}
\frac{1}{g_{k}^{tr}}=k^{2}+a+3\varphi ^{2}+3\sum\nolimits_{k}g_{k}^{tr}\text{%
.}  \label{gap1}
\end{equation}%
One already recognizes in Eq.(\ref{shift1}) the shift equation Eq.(\ref%
{minimshift}) while identifying the width of the gaussian wave function with
the correlator on site $g=g^{tr}\left( x-x\right)
=\sum\nolimits_{k}g_{k}^{tr}$. The second equation commonly called the gap
equation can be rearranged as:%
\begin{equation}
g_{k}^{tr}=\frac{1}{k^{2}+a+3\varphi ^{2}+3g}\text{.}  \label{Gtrnaive}
\end{equation}%
By summing over all momenta $k$ one obtains%
\begin{equation}
g=\sum\nolimits_{k}\frac{1}{k^{2}+a+3\varphi ^{2}+3g}=\frac{1}{2\sqrt{%
a+3\varphi ^{2}+3g}}\text{.}  \label{width}
\end{equation}%
Taking the square of this equation one arrives at Eq.(\ref{minimgap}).

One can see from Eq.(\ref{minimshift}) that the symmetric solution $\varphi
=0$ exists for any $a$ and is given by a root of the cubic equation
\begin{equation}
g_{s}^{3}+\frac{a}{3}g_{s}^{2}-\frac{1}{12}=0\text{,}  \label{symcubiceq}
\end{equation}%
where the subscript $``s"$ stands for ``symmetric". Accordingly the connected
correlator is%
\begin{equation}
g_{k}^{tr}=\frac{1}{k^{2}+m_{s}^{2}}  \label{gtr}
\end{equation}%
with a mass of
\begin{equation}
m_{s}^{2}=a+3g_{s}.  \label{massofS}
\end{equation}%
In the symmetric phase it coincides with the full 2-body correlator and is
given in Figure 2 as a black dashed line.

The would be broken solution can be written as%
\begin{equation}
\varphi ^{2}=-a-3g_{a}\text{,}  \label{asymsol1}
\end{equation}%
with $g_{a}$ satisfying another cubic equation%
\begin{equation}
g_{a}^{3}+\frac{a}{3}g_{a}^{2}+\frac{1}{24}=0\text{.}  \label{asymsol2}
\end{equation}%
Here the subscript $``a"$ stands for ``asymmetric". The solution exists only
in the double well for $a<-(3/2)^{5/3}\approx -1.966$. The correlator in
this case has a mass
\begin{equation}
m_{a}^{2}=-2a-6g_{a}=2\varphi ^{2}\text{.}  \label{asmass}
\end{equation}%
The order parameter correlator takes the form%
\begin{equation}
P^{tr}\left( k\right) =\varphi ^{2}\delta _{k}+\frac{1}{k^{2}+m_{a}^{2}}%
\text{.}  \label{corrasym}
\end{equation}%
This is shown as a brown dashed dot line in Figure 2. One observes that
while the symmetric solution known to be precise at positive $a$, it becomes
much worse at negative $a$ with large absolute value than the one obtained
with apparently erroneous symmetry broken solution.

\subsection{The covariant gaussian approximation}

In the classical approximation truncation of the DS equation means that the
off shell (nonzero $J\left( x\right) $) equation of state is modified. The
higher cumulants are obtained by differentiation of the source with respect
to the shift $\varphi \left( x\right) $. One can try the same strategy
within the gaussian approximation. In the next section, while considering a
more complicated $O\left( 2\right) $ invariant model, we will focus on an
advantage of this approach: it preserves the Ward identities of linearly
represented symmetries of the system. Now we concentrate on technicalities
of the calculation.

For convenience we reprint here the first and second truncated DS equations,
\begin{equation}
J\left( x\right) =\left( -\partial ^{2}+a\right) \varphi \left( x\right)
+\varphi \left( x\right) ^{3}+3\varphi \left( x\right) G^{tr}\left(
x,x\right) \text{,}  \label{shiftoffshell}
\end{equation}%
and
\begin{equation}
\Gamma ^{tr}\left( x,y\right) =\left( -\partial _{x}^{2}+a\right) \delta
\left( x-y\right) +3\delta \left( x-y\right) \varphi \left( x\right)
^{2}+3\delta \left( x-y\right) G^{tr}\left( x,x\right) .
\label{gapeqoffshell}
\end{equation}%
The truncated correlator $G^{tr}\left( x,y\right) $ should be considered
as a functional of $\varphi \left( x\right) $ that is determined by the
above two equations with the condition
\begin{equation}
\int_{y}\Gamma ^{tr}\left( x,y\right) G^{tr}\left( y,z\right) =\delta \left(
x-z\right) .  \label{inverse}
\end{equation}%
As in the case of classical approximation, the ``true" correlator is derived
by taking derivative of the shift equation Eq.(\ref{shiftoffshell}),
\begin{equation}
\Gamma \left( x,y\right) \equiv \frac{\delta }{\delta \varphi \left(
y\right) }J\left( x\right) =\left\{ -\partial ^{2}+a+3\varphi \left(
x\right) ^{2}+3G^{tr}\left( x,x\right) \right\} \delta (x-y)+3\varphi \left(
x\right) \frac{\delta }{\delta \varphi \left( y\right) }G^{tr}\left(
x,x\right) \text{.}  \label{Gammafull}
\end{equation}%
In contrast $\Gamma ^{tr}$ should be viewed as a variational parameter only.
The first term is just$\ \Gamma ^{tr}\left( x,y\right) $, while the second
term is the ``chain correction"\cite{Kovner}. The origin of the notation
comes from the perturbative analysis of the contributions. Denote it by%
\begin{equation}
\frac{\delta }{\delta \varphi \left( z\right) }G^{tr}\left( x,y\right)
\equiv C(z|x,y)\text{,}  \label{chainnotation}
\end{equation}%
and it can be calculated by differentiation of the identity Eq.(\ref{inverse}%
)%
\begin{equation}
\int_{u}\frac{\delta }{\delta \varphi \left( z\right) }\Gamma ^{tr}\left(
x,u\right) G^{tr}\left( u,y\right) +\int_{u}\Gamma ^{tr}\left( x,u\right)
\frac{\delta }{\delta \varphi \left( z\right) }G^{tr}\left( u,y\right) =0%
\text{.}  \label{derivchain}
\end{equation}%
Multiplying from the left by the matrix $G^{tr},$one obtains%
\begin{equation}
C(z|x,y)=-\int_{u,v}G^{tr}(x,v)\frac{\delta \Gamma ^{tr}(v,u)}{\delta
\varphi \left( z\right) }G^{tr}(u,y)\text{.}  \label{derivchain1}
\end{equation}%
Taking derivative of the gap equation, Eq.(\ref{gapeqoffshell}), results in%
\begin{equation}
\frac{\delta \Gamma ^{tr}(v,u)}{\delta \varphi \left( z\right) }=6\varphi
\left( u\right) \delta (z-v)\delta (v-u)+3\delta (v-u)C(z|v,v)\text{.}
\label{derivchain2}
\end{equation}%
Thus the chain equation becomes%
\begin{equation}
C(z|x,y)=-6\varphi \left( z\right)
G^{tr}(x,z)G^{tr}(z,y)-3\int_{v}G^{tr}(x,v)G^{tr}(v,y)C(z|v,v)\text{.}
\label{chaineqx}
\end{equation}%
Note that this equation is linear in the chain function $C$ and can be
conveniently solved by iteration. The problem is that the number of unknowns
is very large. However there are two observations that greatly reduce the
complexity. First the argument $z$ is the same on both the right and left
hand side and therefore is just a parameter. The second is that the right
hand side of the equation depends only on $C(z|v,v)$ with two last arguments
identical. Consequently one can first solve the particular case of $y=x$:%
\begin{equation}
C(z|x,x)=-6\varphi \left( z\right)
G^{tr}(x,z)G^{tr}(z,x)-3\int_{v}G^{tr}(x,v)G^{tr}(v,x)C(z|v,v)\text{.}
\label{particular}
\end{equation}%
This particular chain is in fact the only quantity we need in order to
calculate the ``covariant" correction to the correlator in Eq.(\ref{Gammafull}%
):
\begin{equation}
\Delta \Gamma \left( x,y\right) \equiv \Gamma \left( x,y\right) -\Gamma
^{tr}\left( x,y\right) =3\varphi \left( x\right) C\left( y|x,x\right) \text{.%
}  \label{corrdefagain}
\end{equation}%
We will not need temporarily the chain function for arbitrary arguments.

Using the translation invariance, one sees the function depends on just one
variable:
\begin{equation}
C(z|x,x)=c\left( z-x\right) \text{.}  \label{smallc}
\end{equation}%
It obeys
\begin{equation}
c(z-x)=-6g^{tr}(x-z)g^{tr}(z-x)\varphi -3\int_{v}g^{tr}(x-v)g^{tr}(v-x)c(z-v)%
\text{.}  \label{translationchaineq}
\end{equation}%
In momentum space the linear equation becomes algebraic for one variable
only,%
\begin{equation}
c_{k}=-6\varphi
\sum\nolimits_{p}g_{p}^{tr}g_{k-p}^{tr}-3\sum%
\nolimits_{p}g_{p}^{tr}g_{k-p}^{tr}c_{k}\text{,}  \label{momentumchaineq}
\end{equation}%
which can be easily solved as%
\begin{equation}
c_{k}=-\frac{6\varphi f_{k}}{1+3f_{k}}\text{.}  \label{chainsol}
\end{equation}%
Here the ``fish" diagram is%
\begin{equation}
f_{k}\equiv \sum\nolimits_{p}g_{p}^{tr}g_{k-p}^{tr}=\frac{1}{m_{a}}\frac{1}{%
k^{2}+4m_{a}^{2}}\text{.}  \label{fish}
\end{equation}

Substituting this into the expression for the cumulant correction in the
momentum space, one gets%
\begin{equation}
\Delta \Gamma _{k}=3\varphi c_{k}=-\frac{18\varphi ^{2}}{f_{k}^{-1}+3}\text{.%
}  \label{corrk}
\end{equation}%
The order parameter correlator in CGA finally is%
\begin{equation}
P\left( k\right) =\varphi ^{2}\delta _{k}+\left( k^{2}+m_{a}^{2}-\frac{%
18\varphi ^{2}}{f_{k}^{-1}+3}\right) ^{-1}=\frac{m_{a}^{2}}{2}\delta _{k}+%
\frac{k^{2}+4m_{a}^{2}+\frac{3}{m_{a}}}{\left( k^{2}+M_{1}^{2}\right) \left(
k^{2}+M_{2}^{2}\right) }\text{,}  \label{opcorrelator}
\end{equation}%
where%
\begin{equation}
M_{1,2}^{2}=\frac{1}{2m_{a}}\left( 3+5m_{a}^{3}\pm 3\sqrt{%
1+6m_{a}^{3}+m_{a}^{6}}\right) .  \label{Poles}
\end{equation}%
As an exact statement and reliable simulations, Figure 1, shows there is no
symmetry breaking in this model and thus the appearance of the delta
function is as erroneous as in the classical approximation. Eq.(\ref%
{opcorrelator}) at $k\neq 0$ is presented as a black line in Figure 2,
together with results calculated by different methods in the momentum space$%
. $

Eq.(\ref{opcorrelator}) also demonstrates that there are poles and they do
approximate well with excitations corresponding to the double well bound
states, as one can see in Figure 3. In Figure 4 we show results of a more
complicated invariant correlator $P_{2}\left( k\right) \equiv
\int_{x}e^{-ikx}\left \langle \phi \left( x\right) ^{2}\phi \left( 0\right)
^{2}\right \rangle $ calculated within CGA as well as results of exact
numerical diagonalization. Details of the calculations can be found in
Appendix B.

\begin{figure}[t!]
\centering
\includegraphics[width=8cm]{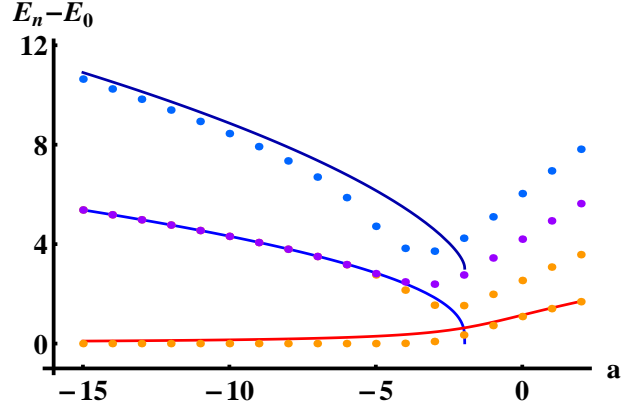}
\caption{Poles and excitations. Dots are excitations $E_{n}-E_{0}$
calculated by exact numerical diagonalization of the double well potential.
The red line is the mass of the two point correlator in symmetric phase Eq.(%
\protect\ref{massofS}). The blue and dark blue lines are poles in the
``broken" phase Eq.(\protect\ref{Poles}) calculated by CGA.}
\label{fig3}
\end{figure}

\begin{figure}[t!]
\centering
\includegraphics[width=14cm]{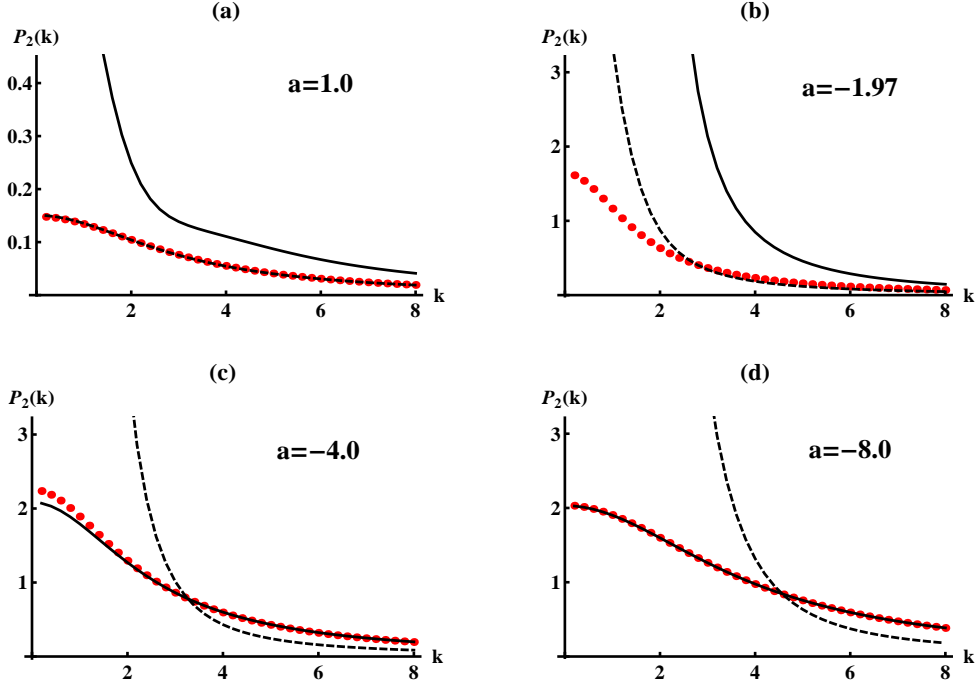}
\caption{Correlator of composite operator $\protect\phi ^{2}\left( r\right) $
at $a=1,-1.97,-4,-8.$ The red dots are exact numerically calculated results.
Black lines are CGA\ results starting from the broken solution of
minimization equations. Black dashed lines are results starting from the
symmetric solution.}
\label{fig4}
\end{figure}

\subsection{Variational interpretation of CGA}

Generally cumulants are collected in the so called effective action\cite%
{Amit}, so that%
\begin{equation}
A_{eff}\left[ \varphi \right] =\sum\limits_{n}\Gamma ^{n}\left(
x_{1},\cdots,x_{n}\right) \varphi \left( x_{1}\right) \cdots \varphi \left(
x_{n}\right).  \label{Aeffdef}
\end{equation}%
The question arises whether the CGA cumulants defined in the previous
subsection can be obtained in the way from an approximated effective action.
The answer is yes. Although due to its complexity the action is of little
use in actual computation and the DS equations truncation method is far
superior. However it will be useful conceptually in the next section to
discuss the symmetry properties of the approximation that are of crucial
importance in applications.

As we have seen already the minimization equations shared by the naive
gaussian approximation and CGA are equivalent to minimization of the
expectation value of Hamiltonian, Eq.(\ref{Hamiltonian}), on a set of
general gaussian wave functions, Eq.(\ref{wavefunction}). The CGA
correlators can be obtained from a truncation of the (in principle exact)
Cornwall-Jackiw-Tomboulis (CJT) functional\cite{CJT} that is a double
Legendre transformation. The action contains infinite number of terms:%
\begin{equation}
A_{eff}\left[ \varphi \left( x\right) ,G\right] =A\left[ \varphi \left(
x\right) \right] +\frac{1}{2}\text{Tr}\left[ \log G^{-1}\right] +\frac{1}{2}%
\text{Tr}\left[ D^{-1}G-1\right] +\frac{3}{2}\int_{x}\varphi \left( x\right)
^{2}G\left( x,x\right) +\frac{3}{4}\int_{x}G\left( x,x\right) ^{2}+\cdots
\text{.} \label{CJT1}
\end{equation}%
Here
\begin{equation}
D^{-1}\left( x,y\right) =\left( -\partial _{x}^{2}+a\right) \delta \left(
x-y\right) \text{,}  \label{invD}
\end{equation}%
and each matrix element of the correlator, $G\left( x,y\right) $, should be
understood as a functional of the order parameter $\varphi \left( x\right) $
determined by the minimization equation $\delta A/\delta G=0$. Traces and
logarithm are also understood in the matrix sense.

Let us truncate this infinite series to the terms explicitly written in Eq.(%
\ref{CJT1}). The minimization condition becomes just the gap equation, Eq.(%
\ref{gapeq}), for the truncated correlator that can be written in the
following form:%
\begin{equation}
-\frac{1}{2}\left( G^{tr}\right) ^{-1}\left( x,y\right) +\frac{1}{2}%
D^{-1}\left( x,y\right) +\frac{3}{2}\varphi \left( x\right) ^{2}\delta
\left( x-y\right) +\frac{3}{2}G^{tr}\left( x,x\right) \delta \left(
x-y\right) \text{,}  \label{CJTgap}
\end{equation}%
and vanish if the gap equation is obeyed. The first functional derivative of
$A_{eff}$ with respect to $\varphi \left( x\right) $ reproduces $J\left(
x\right) $ of the truncated shift equation,%
\begin{equation}
J\left( x\right) =\frac{\delta A_{eff}}{\delta \varphi \left( x\right) }%
=\left( -\partial _{x}^{2}+a\right) \varphi \left( x\right) +\varphi \left(
x\right) ^{3}+3\varphi \left( x\right) G^{tr}\left( x,x\right) \text{,}
\label{CJTshift}
\end{equation}%
that coincides with Eq.(\ref{shifteq}). This means that the second
derivative of the effective action coincides with the CGA correlator rather
than with the gaussian truncated correlator.

\section{Why all the Ward identities are obeyed by the CGA correlators}

\subsection{Importance of symmetry preservation in the phase with broken
symmetry}

In this section we focus on the preservation of all the symmetry properties
of correlators within CGA. These properties are crucially important for
making an approximation useful. The issue is obviously crucial when the
symmetry is spontaneously broken since the low energy properties are
determined by the Goldstone bosons\cite{Lubenskybook} (GB), massless modes
whose very existence hinges on the symmetry breaking. Consequences of the
symmetry whether broken or not in terms of correlators are expressed by the
Ward identities\cite{Amit}. While the classical approximation obeys the Ward
identities, the naive gaussian approximation unfortunately does not.

We demonstrate here that the CGA corrections to the naive gaussian
correlators are just enough to make the full correlators consistent with the
Ward identities. This allows one in particular to take into account
accurately the Goldstone bosons. In addition the ``charge conserving"\cite%
{Baym} property of the approximation is imperative if one discusses
renormalizability (small distance cutoff dependence) with respect to
ultraviolet divergencies. In $d\geq 2$ the reason that the UV cutoff
dependence may be absorbed by the the renormalized wave function and coupling
constants hinges on the symmetry considerations\cite{Amit}.

To this end it will be more instructive to consider a continuous symmetry
rather than the discrete symmetry $Z_{2}$, since we would like to involve
the GB. The model possessing the simplest continuous symmetry, $O\left(
2\right) $, has the following free energy:%
\begin{equation}
A=\frac{1}{2\omega }\int_{\mathbf{r}}\left \{ (\nabla \phi ^{a}\left(
\mathbf{r}\right) )^{2}+a\phi ^{a}\left( \mathbf{r}\right) \phi ^{a}\left(
\mathbf{r}\right) +\frac{1}{2}\left( \phi ^{a}\left( \mathbf{r}\right) \phi
^{a}\left( \mathbf{r}\right) \right) ^{2}\right \} \text{,}
\label{O(2)model}
\end{equation}%
where $a=1,2$. We use two real fields, although very often a complex field
is employed. The dimensionality $d$ will be kept arbitrary for the time
being. The $O(2)$ invariance is expressed in a functional form as
independence of the effective action under the transformation,%
\begin{equation}
A_{eff}\left[ R^{ab}\left( \theta \right) \varphi ^{b}\left( \mathbf{r}%
\right) \right] =A_{eff}\left[ \varphi ^{a}\left( \mathbf{r}\right) \right]
\text{,}  \label{invariance}
\end{equation}%
for any angle $\theta $. Here $R^{ab}\left( \theta \right) $ is the global
symmetry rotation matrix. The infinitesimal transformation, using the chain
rule relates this to the ES:%
\begin{equation}
0=A_{eff}\left[ \varphi ^{a}\left( \mathbf{r}\right) +\theta \varepsilon
^{ab}\varphi ^{b}\left( \mathbf{r}\right) \right] -A_{eff}\left[ \varphi
^{a}\left( \mathbf{r}\right) \right] \Longrightarrow \int_{\mathbf{r}}\frac{%
\delta A_{eff}}{\delta \varphi ^{a}\left( \mathbf{r}\right) }\varepsilon
^{ab}\varphi ^{b}\left( \mathbf{r}\right) =\int_{\mathbf{r}}J^{a}\left(
\mathbf{r}\right) \varepsilon ^{ab}\varphi ^{b}\left( \mathbf{r}\right) =0%
\text{.}  \label{Ward1}
\end{equation}%
Functional derivatives of this equation with respect to $\varphi ^{a}\left(
\mathbf{r}\right) $ generate all the Ward identities expressing the
symmetry. The first two are
\begin{subequations}
\begin{eqnarray}
0 &=&\left. \frac{\delta }{\delta \varphi ^{a}\left( \mathbf{r}\right) }%
\int_{\mathbf{r}^{\prime }}J^{e}\left( \mathbf{r}^{\prime }\right)
\varepsilon ^{ef}\varphi ^{f}\left( \mathbf{r}^{\prime }\right) \right \vert
_{J=0}=\int_{\mathbf{r}^{\prime }}\Gamma ^{ae}\left( \mathbf{r},\mathbf{r}%
^{\prime }\right) \varepsilon ^{ef}\varphi ^{f}\left( \mathbf{r}^{\prime
}\right) ;  \label{WardI2a} \\
0 &=&\left. \frac{\delta ^{2}}{\delta \varphi ^{b}\left( \mathbf{r}%
_{2}\right) \delta \varphi ^{a}\left( \mathbf{r}_{1}\right) }\int_{\mathbf{r}%
^{\prime }}J^{e}\left( \mathbf{r}^{\prime }\right) \varepsilon ^{ef}\varphi
^{f}\left( \mathbf{r}^{\prime }\right) \right \vert _{J=0}  \notag \\
&=&\int_{\mathbf{r}^{\prime }}\Gamma ^{abe}\left( \mathbf{r}_{1},\mathbf{r}%
_{2},\mathbf{r}^{\prime }\right) \varepsilon ^{ef}\varphi ^{f}\left( \mathbf{%
r}^{\prime }\right) +\Gamma ^{ae}\left( \mathbf{r}_{1},\mathbf{r}_{2}\right)
\varepsilon ^{eb}+\Gamma ^{eb}\left( \mathbf{r}_{1},\mathbf{r}_{2}\right)
\varepsilon ^{ea}\text{.}  \label{WardI2b}
\end{eqnarray}
The first equation on shell gives the Goldstone theorem. Indeed taking $%
\varphi ^{1}=\varphi $, $\varphi ^{2}=0$ in momentum space it reads
\end{subequations}
\begin{equation}
\Gamma ^{a2}\left( \mathbf{k=0}\right) =0\text{.}  \label{Goldstonetheorem}
\end{equation}

\subsection{Proof of the conserving property of CGA}

Here we use two quite different arguments to demonstrate that CGA obeys the
Ward identities. The first is to prove Eq.(\ref{Ward1}) directly within a
specific model using the CGA (truncated) off shell ES and the gap equation.
In the $O\left( 2\right) $ model, the ES is,
\begin{equation}
J^{a}\left( \mathbf{r}\right) =\left( -\nabla ^{2}+a\right) \varphi
^{a}\left( \mathbf{r}\right) +\varphi ^{a}\left( \mathbf{r}\right) \varphi
^{e}\left( \mathbf{r}\right) \varphi ^{e}\left( \mathbf{r}\right) +\omega
\varphi ^{a}\left( \mathbf{r}\right) G_{tr}^{ee}\left( \mathbf{r},\mathbf{r}%
\right) +2\omega \varphi ^{e}\left( \mathbf{r}\right) G_{tr}^{ae}\left(
\mathbf{r},\mathbf{r}\right) \text{,}  \label{ESO(2)}
\end{equation}%
and the gap equation takes the form%
\begin{equation}
\Gamma _{tr}^{ab}\left( \mathbf{r}_{1},\mathbf{r}_{2}\right) =\left\{ \delta
^{ab}\left( -\nabla _{r_{1}}^{2}+a+\varphi ^{e}\left( \mathbf{r}_{1}\right)
\varphi ^{e}\left( \mathbf{r}_{1}\right) +\omega G_{tr}^{ee}\left( \mathbf{r}%
_{1},\mathbf{r}_{1}\right) \right) +2\varphi ^{a}\left( \mathbf{r}\right)
\varphi ^{b}\left( \mathbf{r}\right) +2\omega G_{tr}^{ab}\left( \mathbf{r}%
_{1},\mathbf{r}_{1}\right) \right\} \delta \left( \mathbf{r}_{1}-\mathbf{r}%
_{2}\right) \text{.}  \label{gapeqO(2)}
\end{equation}%
Substituting Eq.(\ref{ESO(2)}) into the functional Ward identity, Eq.(\ref%
{Ward1}), one obtains:%
\begin{equation}
\int_{\mathbf{r}}\left\{ -\nabla ^{2}\varphi ^{a}\left( \mathbf{r}\right)
+\left( a+\varphi ^{e}\left( \mathbf{r}\right) \varphi ^{e}\left( \mathbf{r}%
\right) +\omega G_{tr}^{ee}\left( \mathbf{r},\mathbf{r}\right) \right)
\varphi ^{a}\left( \mathbf{r}\right) +2\omega G_{tr}^{ae}\left( \mathbf{r},%
\mathbf{r}\right) \varphi ^{e}\left( \mathbf{r}\right) \right\} \varepsilon
^{ab}\varphi ^{b}\left( \mathbf{r}\right) \text{.}  \label{WIproof}
\end{equation}%
The second term in the curly brackets vanishes identically, while the first
vanishes after integration by parts,%
\begin{equation}
\int_{\mathbf{r}}\varepsilon ^{ab}\varphi ^{b}\left( \mathbf{r}\right)
\nabla ^{2}\varphi ^{a}\left( \mathbf{r}\right) =-\int_{\mathbf{r}%
}\varepsilon ^{ab}\nabla \varphi ^{b}\left( \mathbf{r}\right) \nabla \varphi
^{a}\left( \mathbf{r}\right) =0\text{.}  \label{derterm}
\end{equation}%
The only nontrivial term therefore is the last one,%
\begin{equation}
\int_{\mathbf{r}}\varphi ^{b}\left( \mathbf{r}\right) \varepsilon
^{ba}G_{tr}^{ae}\left( \mathbf{r},\mathbf{r}\right) \varphi ^{e}\left(
\mathbf{r}\right) \text{.}  \label{LasttermWI}
\end{equation}

To show that this term also vanishes, let us multiply the gap equation Eq.(%
\ref{gapeqO(2)}) by $G_{tr}^{bc}\left( \mathbf{r}_{2},\mathbf{r}\right) $
and integrate over $\mathbf{r}_{2}$:
\begin{equation}
\left( -\nabla _{\mathbf{r}_{1}}^{2}+a+\left( \varphi ^{e}\left( \mathbf{r}%
_{1}\right) \right) ^{2}+\omega G_{tr}^{ee}\left( \mathbf{r}_{1},\mathbf{r}%
_{1}\right) \right) G_{tr}^{ac}\left( \mathbf{r}_{1},\mathbf{r}\right)
+2\omega \left( \varphi ^{a}\left( \mathbf{r}_{1}\right) \varphi ^{b}\left(
\mathbf{r}_{1}\right) +G_{tr}^{ab}\left( \mathbf{r}_{1},\mathbf{r}%
_{1}\right) \right) G_{tr}^{bc}\left( \mathbf{r}_{1},\mathbf{r}\right)
=\delta ^{ac}\delta \left( \mathbf{r}_{1}-\mathbf{r}\right) \text{.}
\label{deriv0}
\end{equation}%
Multiply this by $\varepsilon ^{ac}$, summing over $a,c$ and taking $\mathbf{%
r=r}_{1}$ at the end. Thus, the scalar equation simplifies due to symmetry $%
G_{tr}^{ac}\left( \mathbf{r},\mathbf{r}\right) =G_{tr}^{ca}\left( \mathbf{r},%
\mathbf{r}\right) $ into%
\begin{equation}
-\varepsilon ^{ac}\nabla _{\mathbf{r}^{\prime }}^{2}G_{tr}^{ac}\left(
\mathbf{r}^{\prime },\mathbf{r}_{1}\right) |_{\mathbf{r}^{\prime }=\mathbf{r}%
_{1}}+2\varepsilon ^{ac}\varphi ^{a}\left( \mathbf{r}_{1}\right) \varphi
^{b}\left( \mathbf{r}_{1}\right) G_{tr}^{bc}\left( \mathbf{r}_{1},\mathbf{r}%
_{1}\right) +2\omega \varepsilon ^{ac}G_{tr}^{ab}\left( \mathbf{r}_{1},%
\mathbf{r}_{1}\right) G_{tr}^{bc}\left( \mathbf{r}_{1},\mathbf{r}_{1}\right)
=0\text{.}  \label{deriv1}
\end{equation}%
The last term vanishes since $G_{tr}^{ab}\left( \mathbf{r},\mathbf{r}\right)
G_{tr}^{bc}\left( \mathbf{r},\mathbf{r}\right) $ is symmetric under $%
a\leftrightarrow c$. Finally integrating over $\mathbf{r}_{1}$, the first
term vanishes as a full derivative and we arrive at Eq.(\ref{LasttermWI}).

The second method to demonstrate the Ward identities utilizes the known
effective potential within the CGA, given in the previous section, Eq.(\ref%
{CJT1}). A general observation is that it is written covariantly as an $%
O\left( 2\right) $ scalar and in addition the CGA\ off shell minimization
equations are covariant, namely $J^{a}$ and $\varphi ^{a}$ are $O\left(
2\right) $ vectors, while $G_{tr}^{ab}$ is a tensor. This was what
originally motivated the term ``CGA"\cite{Kovner}. Indeed, if $\varphi
^{a}\left( \mathbf{r}\right) $ and $G_{tr}^{ab}\left( \mathbf{r}\right) $
are solutions of the minimization equations Eq.(\ref{ESO(2)}) and Eq.(\ref%
{gapeqO(2)}), so are $R^{ae}\left( \theta \right) \varphi ^{e}\left( \mathbf{%
r}\right) $ and $R^{ae}\left( \theta \right) R^{bf}\left( \theta \right)
G_{tr}^{ef}\left( \mathbf{r}\right) $, provided the source $J\left( \mathbf{r%
}\right) $ was rotated as well. This means that the CGA effective action is
invariant just as the exact one in Eq.(\ref{invariance}) and all the Ward
identities follow. The truncation of the higher correlators doesn't violate
the covariance of $\varphi ^{a}$ and $G_{tr}^{ab}$, and therefore definition
of correlator via derivatives of $J$ keeps all the Ward identities.

The covariance proof of CGA can be extended to any symmetry linearly
represented in the free energy and to statistical physics and (relativistic
or not) many - body system involving fermionic, gauge field as long as the
representation of the symmetry is linear or the truncation is covariant.

\subsection{How it all works in $d=3$}

The proof of Ward identities is rather formal. As an example, let us see
explicitly the $O\left( 2\right) $ invariant model in dimension $d=3$. The
shift and gap equations in momentum space for the asymmetric order parameter
along the direction $1$, $\varphi ^{1}=\varphi $, $\varphi ^{2}=0$, are
\begin{subequations}
\begin{eqnarray}
0 &=&a\varphi +\varphi ^{3}+\omega \varphi \sum\limits_{\mathbf{k}}\left(
G_{tr}^{22}\left( \mathbf{k}\right) +3G_{tr}^{11}\left( \mathbf{k}\right)
\right) ;  \label{shgapmomentum} \\
\Gamma _{tr}^{ab}\left( \mathbf{k}\right) &=&\delta ^{ab}\left(
k^{2}+a+\varphi ^{2}\right) +2\varphi ^{a}\varphi ^{b}+\omega \sum\limits_{%
\mathbf{k}}\left( \delta ^{ab}G_{tr}^{cc}\left( \mathbf{k}\right)
+2G_{tr}^{ab}\left( \mathbf{k}\right) \right) \text{.}
\end{eqnarray}

Due to remaining $Z_{2}$ symmetry $\phi ^{2}\rightarrow -\phi ^{2}$, the
``mixed" correlator $G_{tr}^{12}$ vanishes and the diagonal components of the
cumulant take the form,
\end{subequations}
\begin{equation}
\Gamma _{tr}^{11}=k^{2}+m_{1}^{2};\text{ \ }\Gamma _{tr}^{22}=k^{2}+m_{2}^{2}%
\text{.}  \label{remainingZ2}
\end{equation}%
This leads to the following set of algebraic equations for the two masses
and $\varphi $:
\begin{subequations}
\begin{eqnarray}
0 &=&a+\varphi ^{2}+\frac{3\omega }{2\pi ^{2}}\left( \Lambda -\frac{\pi m_{1}%
}{2}\right) +\frac{\omega }{2\pi ^{2}}\left( \Lambda -\frac{\pi m_{2}}{2}%
\right) ; \\
m_{1}^{2} &=&a+3\varphi ^{2}+\frac{3\omega }{2\pi ^{2}}\left( \Lambda -\frac{%
\pi m_{1}}{2}\right) +\frac{\omega }{2\pi ^{2}}\left( \Lambda -\frac{\pi
m_{2}}{2}\right) ; \\
m_{2}^{2} &=&a+\varphi ^{2}+\frac{\omega }{2\pi ^{2}}\left( \Lambda -\frac{%
\pi m_{1}}{2}\right) +\frac{3\omega }{2\pi ^{2}}\left( \Lambda -\frac{\pi
m_{2}}{2}\right) \text{.}
\end{eqnarray}%
The UV cutoff $\Lambda $ can be absorbed into the renormalized coupling
\end{subequations}
\begin{equation}
a_{r}=a+\frac{2\omega }{\pi ^{2}}\Lambda .  \label{renormA}
\end{equation}%
The solution that exists only for $a_{r}<0.03935\omega ^{2}$ is given in
Appendix C (a symmetric solution, $\varphi =0$, that exists for all values
of $a_{r}>0$ is also given there).

The conclusion is that $G_{tr}^{ab}$ is massive, namely does not have zero
modes. However, as explained in the previous section, the naive gaussian
correlator $G_{tr}^{ab}$ is not the CGA correlator. The later is given by
the derivative of Eq.(\ref{ESO(2)}),
\begin{equation}
\Gamma ^{ab}\left( \mathbf{x},\mathbf{y}\right) =\Gamma _{tr}^{ab}\left(
\mathbf{x,y}\right) +\omega \left[ \varphi ^{a}\left( \mathbf{x}\right)
\frac{\delta }{\delta \varphi ^{b}\left( \mathbf{y}\right) }%
G_{tr}^{ee}\left( \mathbf{x,x}\right) +2\varphi ^{e}\left( \mathbf{x}\right)
\frac{\delta }{\delta \varphi ^{b}\left( \mathbf{y}\right) }%
G_{tr}^{ae}\left( \mathbf{x,x}\right) \right] \text{.}
\label{fullcorrelator}
\end{equation}%
The first term on the right hand side of the above equation is just$\ $the inverse to $%
G_{tr}^{ab}$, while the second term gives the correction $\Delta \Gamma
^{ab}\left( \mathbf{x},\mathbf{y}\right) $ containing the following chain
function
\begin{equation}
\frac{\delta }{\delta \varphi ^{c}\left( \mathbf{z}\right) }%
G_{tr}^{ab}\left( \mathbf{x,y}\right) \equiv C^{c|ab}\left( \mathbf{z|x,y}%
\right) \text{.}  \label{notationchain}
\end{equation}%
It is calculated from the chain equation in momentum space for the relevant
quantity $c_{\mathbf{k}}^{c|ab}$ with the same Fourier transform defined as
in section 2,%
\begin{eqnarray}
&&c_{\mathbf{k}}^{a|bc}+\omega \left( 3\delta _{b1}\delta _{c1}f_{\mathbf{k}%
}^{11}+\delta _{b2}\delta _{c2}f_{\mathbf{k}}^{22}\right) c_{\mathbf{k}%
}^{a|11}+2\omega \left( \delta _{b2}\delta _{c1}+\delta _{b1}\delta
_{c2}\right) f_{\mathbf{k}}^{12}c_{\mathbf{k}}^{a|12}+\omega \left( \delta
_{b1}\delta _{c1}f_{\mathbf{k}}^{11}+3\delta _{b2}\delta _{c2}f_{\mathbf{k}%
}^{22}\right) c_{\mathbf{k}}^{a|22}  \notag \\
&=&-2\varphi \left[ 3\delta _{a1}\delta _{b1}\delta _{c1}f_{\mathbf{k}%
}^{11}+\delta _{a2}\left( \delta _{b1}\delta _{c2}+\delta _{b2}\delta
_{c1}\right) f_{\mathbf{k}}^{12}+\delta _{a1}\delta _{b2}\delta _{c2}f_{%
\mathbf{k}}^{22}\right] .  \label{chainO2}
\end{eqnarray}%
Here the explicit expression for the ``fish" integral,%
\begin{equation}
f_{\mathbf{k}}^{ab}=\sum\nolimits_{\mathbf{p}}g_{tr}^{a}\left( \mathbf{p}%
\right) g_{tr}^{b}\left( \mathbf{k-p}\right) \text{,}  \label{fishO(2)}
\end{equation}%
is listed in Appendix C. The only nonzero components of $c_{\mathbf{k}%
}^{c|ab}$ are:
\begin{subequations}
\begin{eqnarray}
c_{\mathbf{k}}^{1|11} &=&-\frac{2\varphi f_{\mathbf{k}}^{11}(3+8\omega f_{%
\mathbf{k}}^{22})}{1+3\omega \left( f_{\mathbf{k}}^{11}+f_{\mathbf{k}%
}^{22}\right) +8\omega ^{2}f_{\mathbf{k}}^{11}f_{\mathbf{k}}^{22}};
\label{c_results1} \\
c_{\mathbf{k}}^{1|22} &=&-\frac{2\varphi f_{\mathbf{k}}^{22}}{1+3\omega
\left( f_{\mathbf{k}}^{11}+f_{\mathbf{k}}^{22}\right) +8\omega ^{2}f_{%
\mathbf{k}}^{11}f_{\mathbf{k}}^{22}}; \\
c_{\mathbf{k}}^{2|12} &=&c_{\mathbf{k}}^{2|21}=-\frac{2\varphi f_{\mathbf{k}%
}^{12}}{1+2\omega f_{\mathbf{k}}^{12}}\text{.}
\end{eqnarray}%
Substituting this into the expression of the cumulant correction in the
momentum space
\end{subequations}
\begin{equation}
\Delta \Gamma _{\mathbf{k}}^{11}=\omega \varphi \left( 3c_{\mathbf{k}%
}^{1|11}+c_{\mathbf{k}}^{1|22}\right) ;\qquad \Delta \Gamma _{\mathbf{k}%
}^{22}=2\omega \varphi c_{\mathbf{k}}^{2|21}\text{,}  \label{cucorrection}
\end{equation}%
one gets for zero momentum the mass of Goldstone mode:%
\begin{equation}
m_{GB}^{2}=m_{2}^{2}+\Delta \Gamma _{\mathbf{k}=0}^{22}=0\text{.}
\label{GBosons}
\end{equation}%
Therefore the GB reappears in CGA. Similarly more complicated correlators
can be calculated and other Ward identities like Eq.(\ref{WardI2b}) can be
explicitly checked.

\section{Monte Carlo and CGA calculations of the order parameter correlator}

\subsection{ The CGA calculation of the order parameter correlator}

The $O\left( 2\right) $ invariant order parameter correlator in the symmetry
broken phase within CGA is%
\begin{equation}
P\left( \mathbf{k}\right) =\left\langle \phi _{\mathbf{k}}^{a}\phi _{-%
\mathbf{k}}^{a}\right\rangle =\varphi ^{2}\delta _{\mathbf{k}}+G^{11}\left(
\mathbf{k}\right) +G^{22}\left( \mathbf{k}\right) \text{.}  \label{PkO(2)}
\end{equation}%
The contributions for the full CGA expressions containing the correction are
the same as in $d=1$. The explicit expression for $P\left( \mathbf{k}\right)
$ is rather bulky and is not presented here. The integrals similar to those
in 1D were analytically calculated, as shown in Appendix C. For comparison,
the perturbation theory starting from an asymmetric classical solution up to
one loop, is
\begin{equation}
P_{1loop}\left( \mathbf{k}\right) =-a_{r}\delta _{\mathbf{k}}+\omega \left(
\delta _{\mathbf{k}}\frac{3\sqrt{-2a_{r}}}{4\pi }+\frac{1}{k^{2}}+\frac{1}{%
k^{2}-2a_{r}}\right) \text{.}  \label{PoneL}
\end{equation}

The delta function part of order parameter correlator should be included,
when we consider the coordinate space counterpart of Eq.(\ref{PkO(2)}),
\begin{equation}
P\left( \mathbf{r}\right) =\sum\nolimits_{\mathbf{k}}e^{i\mathbf{k\cdot r}%
}P\left( \mathbf{k}\right) =\varphi ^{2}+\sum\nolimits_{\mathbf{k}}e^{i%
\mathbf{k\cdot r}}\left( G^{11}\left( \mathbf{k}\right) +G^{22}\left(
\mathbf{k}\right) \right) \text{.}  \label{Pdiv}
\end{equation}%
It is compared with the one loop result in the coordinate space,%
\begin{equation}
P_{1loop}\left( \mathbf{r}\right) =-a_{r}+\frac{\omega }{4\pi }\left( 3\sqrt{%
-2a_{r}}+\frac{1+e^{-\sqrt{-2a_{r}}\left\vert \mathbf{r}\right\vert }}{%
\left\vert \mathbf{r}\right\vert }\right) \text{,}  \label{PertCorr}
\end{equation}%
in Figure 5, as well as Monte Carlo simulations and perturbation theory on
the lattice.

\begin{figure}[t]
\centering
\includegraphics[width=14cm]{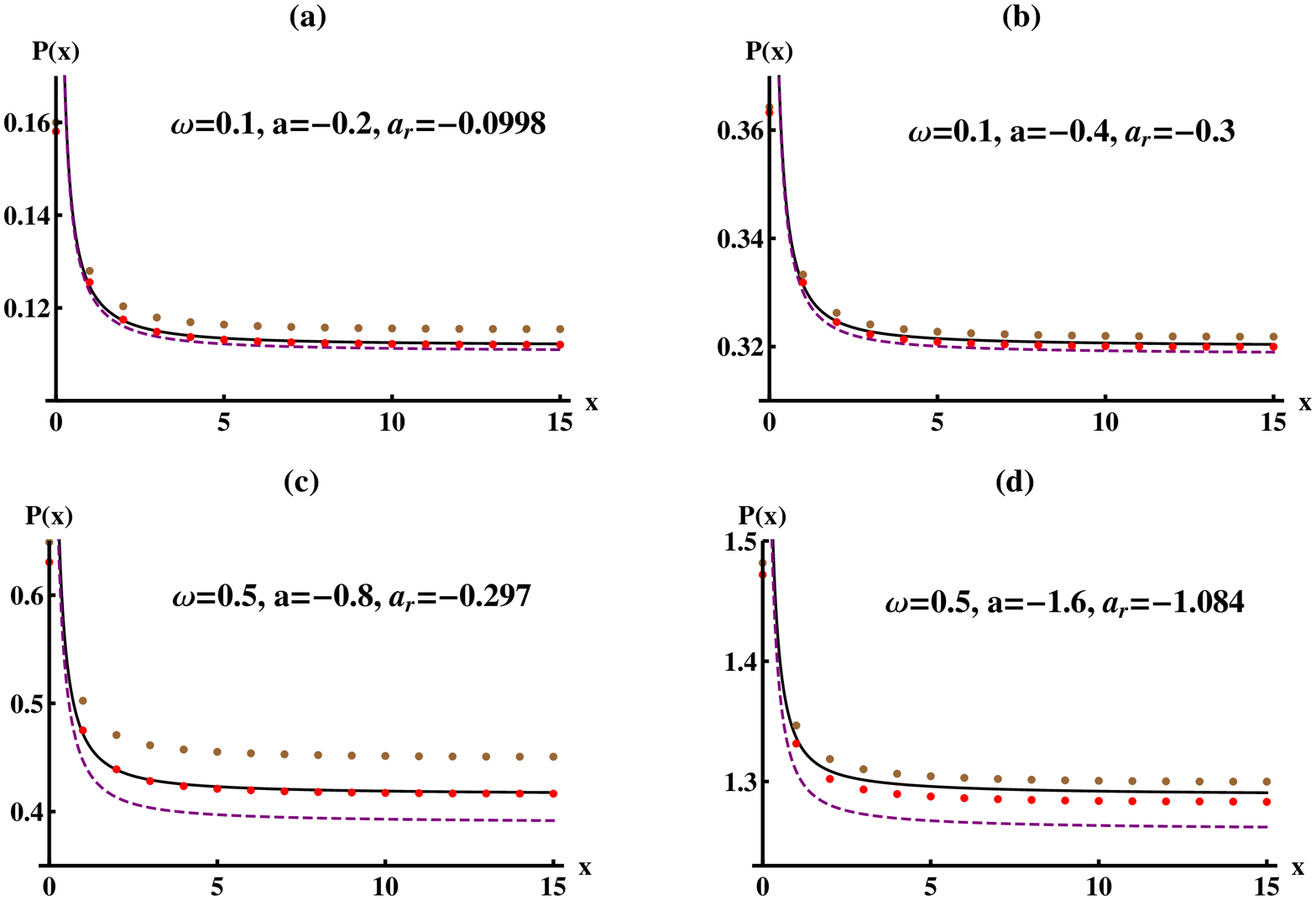}
\caption{Real space comparison for $P(x)$ in the 3D - $O(2)$ model with
different values of $\protect\omega $ and $a$. Black lines are CGA results
Eq.(\protect\ref{Pdiv}), and purple dashed lines are perturbation theory up
to one loop Eq.(\protect\ref{PertCorr}). The brown dots are one loop results
calculated in the lattice model Eq.(\protect\ref{PL}). The red dots are
direct Monte Carlo simulations with error bars smaller than the size of
points. For each bare parameter $a$, the corresponding $a_{r}$ in the
continuum model is obtained by the renormalization condition explained in
section 4.3.}
\label{fig5}
\end{figure}

\subsection{ Monte Carlo simulations of isotropic $\protect\phi ^{4}$ model
in 3D}

To estimate the precision of the CGA, we performed a Monte Carlo simulation
of the scalar model on a cubic lattice of size $N$. The corresponding action
is%
\begin{equation}
S_{L}=\frac{1}{\omega }\sum\limits_{\mathbf{i}}\left[ \frac{1}{2}%
\sum\limits_{\mathbf{\alpha }}\left( \phi _{\mathbf{i}}^{b}-\phi _{\mathbf{%
i-\alpha }}^{b}\right) ^{2}+\frac{a}{2}\phi _{\mathbf{i}}^{b}\phi _{\mathbf{i%
}}^{b}+\frac{1}{4}\left( \phi _{\mathbf{i}}^{b}\phi _{\mathbf{i}}^{b}\right)
^{2}\right] ,  \label{ActionL}
\end{equation}%
where $\mathbf{i}$ is the discrete position and $\mathbf{\alpha }$\textbf{\ }%
is the unit vector along the three axes. With $O(2)$ symmetry, the summation
over $b=1,2$ is implicitly assumed. This model has already been precisely
simulated by several groups\cite{Arnold,Hasenbusch} that focused mainly on
the critical region. Our interest here is to calculate various quantities in
the region of relatively strong thermal fluctuations below the critical
point.

The standard Metropolis algorithm is usually inefficient in the broken phase
because of the large autocorrelation of the samples. The autocorrelation
however can be reduced to a large extent by combining the Metropolis
algorithm with the cluster algorithm. This is done by embedding the Ising
variables into the $\Psi ^{4}$ model with $O(2)$ symmetry and using Wolff's
single-cluster flipping method\cite{Brower,Wolff}. Each cycle of the Monte
Carlo iteration contains a single cluster update of the embedded Ising
variables, followed by a sweep of local updates of the original fields $\phi
^{1}$ and $\phi ^{2}$ using Metropolis algorithm. The method was first
tested in the free model, that is, without the $\phi ^{4}$ terms, for a
small sample size $N=14$. The calculated integrated autocorrelation time was
typically less then $10$. With such reduced autocorrelation, the statistical
error for a run containing several $10^{5}$ cycles after reaching
equilibrium is already small enough.

\subsection{Comparison with CGA}

For measurements sample size of $N=30$ with periodic boundary condition is
used. In order to compare with analytic calculations, the bare parameter $a$
used in the lattice model Eq.(\ref{ActionL}) has to be related to the
renormalized one, $a_{r}$, in Eq.(\ref{renormA}). The UV cutoff there is,
roughly speaking, proportional to the inverse of the lattice distance in a
discrete model. While a better way to relate the two parameters is to
calculate the same quantity both in the continuum and discrete model, and
then compare the results. To this end we use perturbation theory on the
lattice. The two point correlator calculated on the lattice,%
\begin{equation}
P_{\mathbf{i}}^{L}=-a+\frac{\omega }{N^{3}}\sum\limits_{\mathbf{k}}\left(
\frac{e^{i\frac{2\pi }{N}\mathbf{k\cdot i}}-3}{-2a+4\sum_{i}\sin ^{2}\frac{%
\pi k_{i}}{N}}+\frac{e^{i\frac{2\pi }{N}\mathbf{k\cdot i}}-1}{4\sum_{i}\sin
^{2}\frac{\pi k_{i}}{N}}\right) \text{,}  \label{PL}
\end{equation}%
is equated up to order $\omega $ with the same quantity calculated in the
continuum model Eq.(\ref{PertCorr}) at a particular position, like $\left(
N/2,0,0\right) .$

In Figure 5 the Monte Carlo simulations of the order parameter correlator
for $\omega =0.1$ (relatively small fluctuations) and $0.5$ (relatively
strong fluctuations) are presented. Two different values of $a$ in the
broken phase are given in each case. The red points are MC results, and the
black lines are CGA results. The brown dots and purple dashed lines are
results of perturbation theory in the lattice and continuum model
respectively. One observes that the CGA is much closer to the MC simulations
than the perturbation theory in the case of strong fluctuations, which is
relevant to high $T_{c}$ superconductors considered in the next section.

\section{Comparison with experiments on penetration depth of high $T_{c}$
superconductors and discussion}

In this section we employ the CGA method to calculate the magnetic
penetration depth of a type-II superconductor and compare it with both MC
simulations and experiments. Let us first recall the derivation of the
magnetic penetration depth within the GL approach.

\subsection{Penetration depth of a strongly fluctuating superconductor}

Fluctuating magnetic field minimally coupled to the order parameter of a
superconductor (represented by a complex field, $\Psi \left( \mathbf{r}%
\right) =\phi ^{1}\left( \mathbf{r}\right) +i\phi ^{2}\left( \mathbf{r}%
\right) $, in the present context) is contained in the following anisotropic
3D Ginzburg Landau model:
\begin{equation}
F_{GL}=\int_{\mathbf{r}}\left \{ \frac{\hbar ^{2}}{2m^{\ast }}\left \vert
\mathbf{D}_{xy}\Psi \right \vert ^{2}+\frac{\hbar ^{2}}{2m_{c}^{\ast }}\left
\vert D_{z}\Psi \right \vert ^{2}+\alpha \left( T-T_{mf}\right) \left \vert
\Psi \right \vert ^{2}+\frac{b}{2}\left \vert \Psi \right \vert ^{4}+\frac{%
\left( \mathbf{\nabla \times A}\right) ^{2}}{8\pi }\right \} \text{.}
\label{GinzburgLandau}
\end{equation}%
Here $\mathbf{D}\equiv \mathbf{\nabla }+i2\pi \mathbf{A}/\Phi _{0}$ is the
covariant derivative and $T_{mf}$ is the mean - field phase transition
temperature. Here we have assumed that the (bare) coefficient of $%
\left
\vert \Psi \right \vert ^{2}$ is linear in temperature, while other
coefficients are temperature independent. This approximation is reasonable
in a rather wide range of temperatures around $T_{c}$ for high temperature
superconductors\cite{RMP}.

There are two basic scales, the magnetic penetration depth and the coherence
length.\ The zero temperature magnetic penetration depth is $\lambda
_{0}^{2}=\Phi _{0}^{2}m^{\ast }b/(4\pi h^{2}\alpha T_{mf})$ and together
with the $ab$-plane coherence length, $\xi ^{2}=\hbar ^{2}/(2m^{\ast }\alpha
T_{mf})$, it defines a dimensionless parameter\ $\kappa =\lambda _{0}/\xi $,
which is much larger than unity for a typical type-II superconductors. In
anisotropic superconductors, like the high $T_{c}$ cuprates, the ratio $%
\gamma ^{2}=m_{c}^{\ast }/m^{\ast }$ is large and an additional coherence
length scale along the $c$ axis, $\xi _{c}^{2}=\hbar ^{2}/(2m_{c}^{\ast
}\alpha T_{mf})=\xi ^{2}/\gamma ^{2}$, appears. After a scaling of $%
(x,y,z)\rightarrow (\xi x,\xi y,\xi _{c}z)$\textbf{, }$(A_{x},A_{y},A_{z})%
\rightarrow (\Phi _{0}/2\pi \kappa )(A_{x}/\xi ,A_{y}/\xi ,A_{z}/\xi _{c})$
and$\ \Psi \rightarrow (T_{mf}\alpha /b)^{1/2}\Psi $, one writes the
Boltzmann factor at temperature $T$ as
\begin{equation}
S\left[ \Psi ,\mathbf{A}\right] =\frac{1}{2\omega }\int_{\mathbf{r}}\left \{
\left \vert \left( \nabla +i\kappa ^{-1}\mathbf{A}\right) \Psi \right \vert
^{2}+a \left \vert \Psi \right \vert ^{2}+\frac{1}{2}\left \vert \Psi \right
\vert ^{4}\right \} +S_{M}\left[ \mathbf{A}\right],  \label{Boltzman}
\end{equation}%
where%
\begin{equation}
S_{M}\left[ \mathbf{A}\right] =\frac{1}{2\omega }\int_{\mathbf{r}}\left \{
\gamma ^{2}\left( \mathbf{\nabla \times A}\right) ^{2}+\left( 1-\gamma
^{2}\right) \left( \mathbf{\partial }_{x}A_{y}-\mathbf{\partial }%
_{y}A_{x}\right) ^{2}\right \} \text{.}  \label{MWterm}
\end{equation}%
Here two dimensionless parameters
\begin{eqnarray}
\omega &=&\frac{bT}{2\alpha ^{2}T_{mf}^{2}\xi ^{2}\xi _{c}}  \label{omega} \\
a&=&\frac{T}{T_{mf}}-1  \label{aofT}
\end{eqnarray}
were introduced. One observes that complexity of the anisotropy is shifted
to the kinetic term of the vector field.

To derive mesoscopically the macroscopic electrodynamics of the
fluctuating superconductor including the $ab$-plane magnetic penetration
depth (or, in the language of field theory the inverse photon mass), one
expands the effective action of the photon field to the order of $A^{2},$
\begin{equation}
S_{ind}^{eff}[\mathbf{A}]=\frac{1}{2\omega \kappa ^{2}}\sum\limits_{\mathbf{k%
}}A_{\mathbf{k}}^{\alpha }A_{-\mathbf{k}}^{\beta }\left[ \delta _{\alpha
\beta }\sum\nolimits_{\mathbf{q}}\left\langle \left\vert \Psi _{\mathbf{q}%
}\right\vert ^{2}\right\rangle -\frac{1}{\omega }\left\langle J_{\mathbf{k}%
}^{\alpha }J_{-\mathbf{k}}^{\beta }\right\rangle _{c}\right] \text{,}
\label{Seff}
\end{equation}%
and then calculates the averages over the order parameter field $\Psi $
within the $\Psi ^{4}$ isotropic GL model as in the previous sections. For
large $\kappa $ the expression coincides with the leading order in expansion
in $\kappa ^{-2}$. Here
\begin{equation}
J_{\mathbf{k}}^{\alpha }=\frac{1}{2}\sum_{\mathbf{p}}\left( 2\mathbf{p+k}%
\right) _{\alpha }\Psi _{\mathbf{p}}^{\ast }\Psi _{\mathbf{p+k}}\text{,}
\label{current}
\end{equation}%
is the Noether current density of the $O\left( 2\right) $ symmetry defined
in the pure $\Psi ^{4}$ model.

Its connected correlator can be decomposed into the transversal and the
longitudinal parts,%
\begin{equation}
\left \langle J_{\mathbf{k}}^{\alpha }J_{-\mathbf{k}}^{\beta }\right \rangle
_{c}\equiv \left \langle J_{\mathbf{k}}^{\alpha }J_{-\mathbf{k}}^{\beta
}\right \rangle -\left \langle J_{\mathbf{k}}^{\alpha }\right \rangle \left
\langle J_{-\mathbf{k}}^{\beta }\right \rangle =\chi _{T}\left( \delta
_{\alpha \beta }-\frac{k_{\alpha }k_{\beta }}{k^{2}}\right) +\chi _{L}\frac{%
k_{\alpha }k_{\beta }}{k^{2}}\text{.}  \label{connectedJJ}
\end{equation}%
The corresponding coefficient functions, $\chi _{T}$ and $\chi _{L}$, depend
on $k^{2}$ only. The term $\sum \nolimits_{\mathbf{q}}\left \langle |\Psi _{%
\mathbf{q}}|^{2}\right \rangle $ in the effective action Eq.(\ref{Seff}) is
equal to $\chi _{L}/\omega $ due to the ``Ward identity" (derived in Appendix
D). With this replacement, the induced Boltzmann factor is manifestly gauge
invariant and transversal,%
\begin{equation}
S_{ind}^{eff}[\mathbf{A}]=\frac{1}{2\omega ^{2}\kappa ^{2}}\sum \limits_{%
\mathbf{k}}A_{\mathbf{k}}^{\alpha }A_{-\mathbf{k}}^{\beta }\left( \chi
_{L}(k^{2})-\chi _{T}(k^{2})\right) \left( \delta _{\alpha \beta }-\frac{%
k_{\alpha }k_{\beta }}{k^{2}}\right) .  \label{induced}
\end{equation}

The magnetic penetration depth for magnetic field along the $z$ direction, $%
B_{k}^{z}=ik_{x}A_{k}^{y}-ik_{y}A_{k}^{x}$, is now derived through the
classical equation of motion, i.e. taking derivatives of the effective
action with respect to $A_{\mathbf{k}}^{\mu }$. Using the Coulomb gauge, $%
k_{\alpha }A_{\mathbf{k}}^{\alpha }=0$, one obtains%
\begin{equation}
0=\left[ k_{x}^{2}+k_{y}^{2}+\gamma ^{2}k_{z}^{2}+\frac{\chi
_{L}(k^{2})-\chi _{T}(k^{2})}{\omega \kappa ^{2}}\right] B_{k}^{z}\text{.}
\label{equationBz}
\end{equation}%
Therefore the $ab$-plane AC penetration depth in a homogeneous relatively
large sample is
\begin{equation}
\frac{\kappa ^{2}}{\lambda ^{2}}=\lim_{k\rightarrow 0}\frac{\chi
_{L}(k^{2})-\chi _{T}(k^{2})}{\omega }.  \label{penetrationDepth}
\end{equation}%
The remaining work is to calculate the above quantity within the global $%
O(2) $ GL model.

\subsection{CGA calculation of magnetic penetration depth}

In this subsection we calculate the penetration depth Eq.(\ref%
{penetrationDepth}) using CGA. First let us decompose the order parameter
into its real and imaginary parts $\Psi \left( \mathbf{r}\right) =\phi
^{1}\left( \mathbf{r}\right) +i\phi ^{2}\left( \mathbf{r}\right) $, so that
all the quantities are calculated with CGA in the 3D $O(2)$ model as in
previous sections. As explained above the only quantity needed is the
current-current correlator that using real fields takes the form:
\begin{equation}
\left\langle J_{\mathbf{k}}^{\alpha }J_{-\mathbf{k}}^{\beta }\right\rangle
_{c}=-\sum_{\mathbf{pq}}\left( 2\mathbf{p}-\mathbf{k}\right) _{\alpha
}\left( 2\mathbf{q+k}\right) _{\beta }\left\langle \phi _{\mathbf{k-p}%
}^{1}\phi _{\mathbf{p}}^{2}|\phi _{-\mathbf{k-q}}^{1}\phi _{\mathbf{q}%
}^{2}\right\rangle _{c}.  \label{Jviafi}
\end{equation}%
The notation $\left\langle AB|CD\right\rangle _{c}$ in the above equation
means one of $A,B$ should be connected to one of $C,D$. This quantity
therefore contains terms of three and four point connected correlators like $%
\left\langle \phi _{\mathbf{k-p}}^{1}\phi _{\mathbf{p}}^{2}\phi _{-\mathbf{%
k-q}}^{1}\right\rangle _{c}$ and $\left\langle \phi _{\mathbf{k-p}}^{1}\phi
_{\mathbf{p}}^{2}\phi _{-\mathbf{k-q}}^{1}\phi _{\mathbf{q}%
}^{2}\right\rangle _{c}$. Calculation of these terms requires high order
cumulants and within CGA is a combination of the ``triangle" and ``box"
integrals of gaussian truncated Green functions (for an example, see
Appendix B, where it was given for the 1D $Z_{2}$ model). The calculations
of these terms are cumbersome in the 3D case. On the other hand they are
believed to give rise to only high order corrections. Therefore we simply
ignore all these contributions and approximate the current correlator as
\begin{equation}
\left\langle J_{\mathbf{k}}^{\alpha }J_{-\mathbf{k}}^{\beta }\right\rangle
_{c}=\varphi ^{2}k_{\alpha }k_{\beta }G_{\mathbf{k}}^{22}+\sum\nolimits_{%
\mathbf{p}}\left( 2\mathbf{p}-\mathbf{k}\right) _{\alpha }\left( 2\mathbf{p-k%
}\right) _{\beta }G_{\mathbf{k}-\mathbf{p}}^{11}G_{\mathbf{p}}^{22}\text{,}
\label{currentCorrAppr}
\end{equation}%
where $\left\langle \phi _{\mathbf{k}}^{1}\right\rangle =\varphi \delta _{%
\mathbf{k}},\left\langle \phi _{\mathbf{k}}^{2}\right\rangle =0$ and $G_{%
\mathbf{k}}^{ab}=\left\langle \phi _{\mathbf{k}}^{a}\phi _{-\mathbf{k}%
}^{b}\right\rangle _{c}$. With this approximation the penetration depth is
given by
\begin{equation}
\frac{\kappa ^{2}}{\lambda ^{2}}=\frac{1}{\omega }\lim_{k\rightarrow
0}\left( \frac{3k_{\alpha }k_{\beta }}{2k^{2}}-\frac{\delta _{\alpha \beta }%
}{2}\right) \left\langle J_{\mathbf{k}}^{\alpha }J_{-\mathbf{k}}^{\beta
}\right\rangle _{c}=\frac{1}{\omega }\lim_{k\rightarrow 0}\left[ \varphi
^{2}k^{2}G_{\mathbf{k}}^{22}+\sum_{\mathbf{p}}\left( k^{2}-4\mathbf{k\cdot p+%
}\frac{6\left( \mathbf{k\cdot p}\right) ^{2}}{k^{2}}-2p^{2}\right) G_{%
\mathbf{k}-\mathbf{p}}^{11}G_{\mathbf{p}}^{22}\right] ,  \label{pd2}
\end{equation}%
where the momentum integral in the above equation vanishes in the small $k$
limit.

The magnetic penetration depth calculated this way is therefore proportional
to $\varphi ^{2}$. Using reasonable values of $\omega $ and $a_{r}$ the
above result is in good agreement with MC simulation and the experimental
results on various high temperature superconductors.

\subsection{Monte Carlo simulation of the penetration depth. Comparison with
CGA}

The current-current correlator is simulated in the coordinate space for
different values of $\omega $ and $a$. In order to compare with the CGA
results in the momentum space Eq.(\ref{pd2}), finite Fourier transform of
the correlators on the lattice is performed. The penetration depth, or the
superfluid density, at finite momentum%
\begin{equation}
\rho _{s}(k)\equiv \frac{1}{\omega }\left( \chi _{L}(k^{2})-\chi
_{T}(k^{2})\right)  \label{finitK}
\end{equation}%
is then extracted from the simulated current correlator.

In Figure 6 results of $\rho _{s}(k)$ at small but finite wave vectors for $%
\omega =0.1$ and $0.5$ are presented respectively. Red dots with error bars
are MC results while black lines are CGA results. Except for the
discrepancies at large $k,$ or equivalently small distance between the
lattice and continuum model, within the statistical error the two results in
small $k$ limit fit well to each other.

\begin{figure}[t]
\centering
\includegraphics[width=16cm]{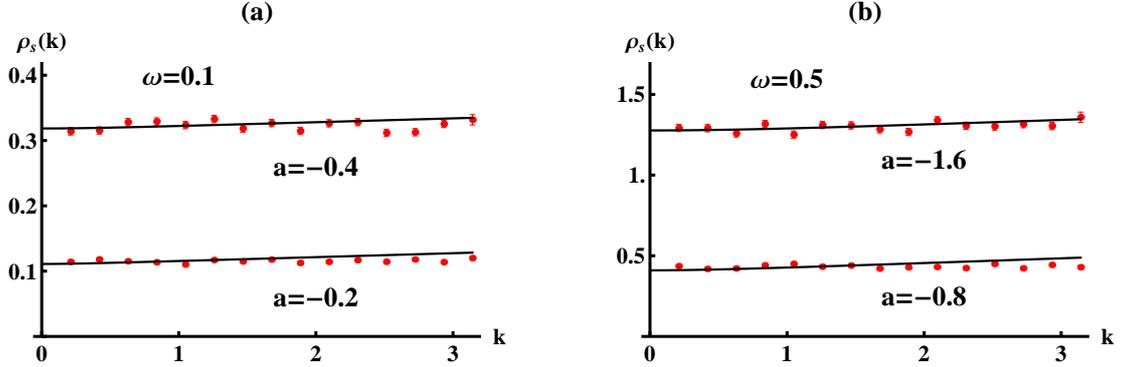}
\caption{Comparison of finite momentum penetration depth $\protect\rho %
_{s}(k)$ for (a) $\protect\omega =0.1$ and (b) $\protect\omega =0.5$ between
CGA\ and Monte Carlo calculations. Black lines are CGA results and red dots
with error bars are MC results. For each parameter $a$, the corresponding $%
a_{r}$ for the continuum model is the same as that in Figure 5.}
\label{fig6}
\end{figure}

\subsection{Comparison with experiments on penetration depth of high $T_{c}$
superconductors}

There exists a large amount of experimental data on temperature dependence
of magnetic penetration depth in high $T_{c}$ superconductors. Most
effective experimental methods include the microwave surface impedance
measurement\cite{Shih-Fu Lee,Broun} and the two-coil mutual inductance
technique\cite{Yuri,Fiory}. Both of these two methods determine indirectly
the microwave conductivity, $\sigma =$ $\sigma ^{\prime }+i\sigma ^{\prime
\prime }$. The superfluid density $\rho _{s}(T)\propto \lambda ^{-2}(T)$ is
then extracted from the imaginary part of the conductivity $\sigma ^{\prime
\prime }$.

In Figure 7a, b, c we compare our CGA results (black lines) with experiments
on three different strongly fluctuating high $T_{c}$ cuprates,
optimally-doped bulk materials\cite{Shih-Fu Lee,Broun} $%
Bi_{2}Sr_{2}CaCu_{2}O_{8}$, $Tl_{2}Ba_{2}CuO_{6+y}$, and $%
YBa_{2}Cu_{3}O_{7-\delta }$ thin film with a thickness of 10 unit cells\cite%
{Yuri}, respectively. The dimensionless ratio, $\lambda _{ab}^{2}(0)/\lambda
_{ab}^{2}(T)$, is shown within about one third of the whole temperature
range below $T_{c}$, where the GL model with linear temperature dependence
of the bare coefficient $a$ Eq.(\ref{aofT}) is still applicable. Straight
dashed lines in Figure 7 are mean-field approximation of the corresponding
material. At lower temperatures it is tangential to both CGA and
experimental data. The estimated mean field critical temperature $T_{mf}$ is
the intersection point between the dashed line and the $T$ axis. One
observes that renormalization of $T_{mf}$ to $T_{c}$ given in gaussian
approximation by Eq.(\ref{renormA})%
\begin{equation}
T_{c}=T_{mf}\left( 1-\frac{2\omega T_{mf}\Lambda \xi }{\pi ^{2}T}\right) ,
\label{Tc}
\end{equation}%
compares reasonably well with experiments for value of UV\ momentum cutoff
of order $\Lambda =0.25/\xi $, deduced from the fluctuation diamagnetism\cite%
{Jiang} for somewhat different materials.

Another feature is the downward curvature of the inverse magnetic
penetration depth within a temperature range below the critical point that
is much wider than the critical region. This is possibly a non - universal
phenomenon caused by strong thermal fluctuations. Given experimentally
measured critical temperature $T_{c}$, the dimensionless thermal fluctuation
strength parameter $\omega $ ($\omega T_{c}/T$, proportional to the square
root of Ginzburg number $Gi$) is the only parameter used to fit the
experimental data. For $Bi_{2}Sr_{2}CaCu_{2}O_{8}$ the resulting value is $%
\omega =1$ (corresponding to $Gi=0.013$), while for $YBa_{2}Cu_{3}O_{7-%
\delta }$\ and $Tl_{2}Ba_{2}CuO_{6+y}$ one gets $\omega =$ $0.5$ ($Gi=0.0032$%
) and $\omega =0.54$ ($Gi=0.0037$) respectively.

One observes that the downward cusp, while absent in the mean field result
is described reasonably well by CGA. Note however that all the three
materials are marginally three dimensional, and the use of the anisotropic
3D GL, Eq.(\ref{GinzburgLandau}), for these highly anisotropic materials is
justified since the coherence length in the $c$ direction (perpendicular to
the $CuO$ planes) exceeds the layer spacing. Perhaps the Lawrence - Doniach
model can give an improved description. Width of the sample even for the 10
unit cells (each containing a bilayer of the $CuO$ planes) $YBCO$ is still
large enough to neglect the finite size effect. It is not clear whether the
Kosterlitz - Thouless transition takes place. Generalizing the CGA method to
2D or layered superconductors to describe the 2D-3D crossover is beyond the
scope of the present paper.

\begin{figure}[t]
\centering
\includegraphics[width=16cm]{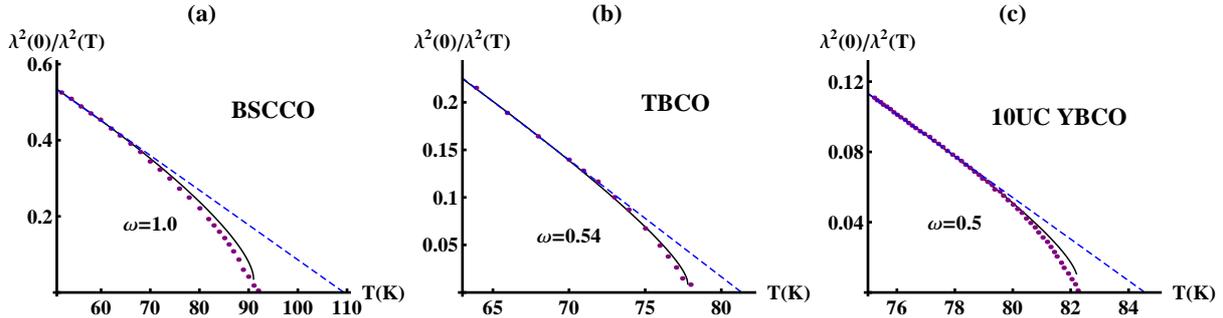}
\caption{Comparison of experimentally measured magnetic penetration depth in
(a) BSCCO, (b) TBCO, and (c) 10 unit cells YBCO with CGA calculations. The
purple dots are experimental data. The black lines are CGA results. The blue
dashed lines are the corresponding mean field behaviors.}
\label{fig7}
\end{figure}

\section{Conclusions}

To summarize, we have developed a non - perturbative method to account for
the strong thermal fluctuations within phenomenological Ginzburg - Landau
approach. The approximation can be broadly described as a systematic
correction of the Hartree - Fock type of mean field description of condensed
matter systems undergoing second order transition. The correction to any
correlator (one or two - body considered in the paper) makes the
approximation ``covariant", i. e. it obeys all the Ward identities of the
relevant symmetry. The development of such an improvement scheme is
motivated by recent experimental realization (in magnetization, conductivity
and Nernst effect) that the fluctuation effects are strong in a much wider
region both above and below $T_{c}$ than the narrow (even in high $T_{c}$
superconductors) critical fluctuation region (determined by the Ginzburg
criterion, $\left\vert T/T_{c}-1\right\vert \ll Gi$), and the theoretical
requirement of a conserving approximation for calculating quantities that hinge
on the symmetry like the magnetic penetration depth in superconductors.

We have demonstrated how all the physical consequences of the symmetry like
the Goldstone theorem and gauge invariance of the current correlator (that
enters the calculation of the magnetic penetration depth) are restored in
the CGA. The method was tested on solvable one - dimensional models and by
comparison with direct Monte Carlo simulation of realistic 3D model. It
turns out that the covariant gaussian approximation captures correctly the
excitation branches in addition to the modes described by the mean field
approximation. This is clearly demonstrated by calculation of the four -
field correlators in a toy model.

We have performed the Monte Carlo simulations of the magnetic penetration
depth in the symmetry broken phase of the 3D Ginzburg - Landau model. It
compares well with CGA in the range accessible for the MC evaluation. The
experimental measurements of the temperature dependence of penetration depth
in high $T_{c}$ cuprates including the downward curvature induced by strong
fluctuations is well captured by the CGA calculations.

Recently it has been demonstrated that several new monolayer 2D materials
like $FeSe$ on $STO$ substrate\cite{Xue} to be superconducting. The high
critical temperature and low dimensionality ensures strong thermal
fluctuations. The corresponding superconducting transition of these 2D
materials can be of the Kosterlitz-Thouless type\cite{Lubenskybook}. It
would be interesting to apply the CGA\ approach to the two - dimensional GL
model. It is not straightforward to describe KT phase transitions, since the
$O\left( 2\right) $ symmetric toy model considered in the present paper
demonstrates there are infrared divergencies\cite{Kao}, and it will be
considered in a later work. The method can be generalized to time dependent
Ginzburg - Landau equations and to the many - body system in which quantum
fluctuations are included on the mesoscopic scale. Then the discussion of
the fluctuations effects in transport can be quantitatively addressed.
Moreover, it is well known that strong magnetic field enhances the thermal
fluctuations, and magnetic field and vortex physics can also be easily
incorporated using the CGA.

\section*{Acknowledgment}

Authors are very grateful to B. Shapiro and R.C Ma for numerous discussions. H.
Kao and B. Rosenstein were supported by MOST of Taiwan through Contract
Grant 104-2112-M-003-012, and 103-2112-M-009-009-MY3. J.F. Wang and D.P. Li were
supported by National Natural Science Foundation of China (No. 11274018 and No. 11674007).
B.R. is grateful to School of Physics of Peking University and Bar Ilan
Center for Superconductivity for hospitality.

\appendix

\section{Calculation of the invariant correlators in d=1 (Quantum Mechanics)}

We use the numerical diagonalization of the quantum mechanics to compute the
correlators. The correlator after renaming $\phi \left( x\right) $ of
statistical physics by $x\left( t\right) $ of Euclidean quantum mechanics ($t
$ being the Matsubara time) is%
\begin{equation}
P\left( t\right) =\Theta \left( t\right) \left\langle 0\left\vert
e^{tH}xe^{-tH}x\right\vert 0\right\rangle +\Theta \left( -t\right)
\left\langle 0\left\vert xe^{tH}xe^{-tH}\right\vert 0\right\rangle \text{,}
\label{A1}
\end{equation}%
where $\Theta \left( t\right) $ is the step function and the Hamiltonian of
the double well is%
\begin{equation}
H=-\frac{1}{2}\partial ^{2}+\frac{a}{2}x^{2}+\frac{1}{4}x^{4}\text{.}
\label{A2}
\end{equation}%
Sandwiching the full set of eigenstates, one obtains%
\begin{eqnarray}
P\left( t\right)  &=&\sum\nolimits_{n>0}\left\vert \left\langle 0\left\vert
x\right\vert n\right\rangle \right\vert ^{2}\left\{ \Theta \left( t\right)
e^{-\left( E_{n}-E_{0}\right) t}+\Theta \left( -t\right) e^{\left(
E_{n}-E_{0}\right) t}\right\} ;  \label{A3} \\
P\left( \omega \right)  &=&\left\vert \left\langle 0\left\vert x\right\vert
0\right\rangle \right\vert ^{2}2\pi \delta \left( \omega \right)
+\sum\nolimits_{n>0}\left\vert \left\langle 0\left\vert x\right\vert
n\right\rangle \right\vert ^{2}\frac{2\left( E_{n}-E_{0}\right) }{\omega
^{2}+\left( E_{n}-E_{0}\right) ^{2}}\text{.}
\end{eqnarray}%
In particular $P\left( \omega =0\right) =2\left\langle \left\vert x\frac{1}{%
H-E_{0}}x\right\vert \right\rangle $. These may be easily calculated
numerically and presented in Figure 1 (as a function of Matsubara time) and
Figure 2 (as a function of frequency).

\section{Correlator of composite operator $\protect\phi ^{2}\left( r\right) $
in the 1D $Z_{2}$ model}

In section 2 we have derived the two - point cumulant $\Gamma \left(
x,y\right) $ within the CGA by taking derivative of the shift equation with
respect to $\varphi \left( r\right) .$ Similarly, one can derive higher
order cumulants by taking more and more derivatives, for example, the three
- point cumulant is (again we set $\omega =1$),
\begin{equation}
\Gamma _{3}\left( z,x,y\right) =\frac{\delta \Gamma \left( x,y\right) }{%
\delta \varphi \left( z\right) }=3\left[ 2\varphi \left( x\right) \delta
(x-y)+C(y|x,x)\right] \delta (x-z)+3C(z|x,x)\delta (x-y)+3\varphi \left(
x\right) C_{4}(y,z|x,x)\text{.} \label{C1}
\end{equation}%
Therefore the function,
\begin{equation}
C_{4}(y,z|x,x^{\prime })\equiv \frac{\delta ^{2}G^{tr}\left( x,x^{\prime
}\right) }{\delta \varphi \left( z\right) \delta \varphi \left( y\right) }=%
\frac{\delta C(y|x,x^{\prime })}{\delta \varphi \left( z\right) },
\label{C2}
\end{equation}%
is the only new unknown chain function that we have to calculate. Performing
the functional derivative of Eq.(\ref{derivchain1}), one obtains:%
\begin{eqnarray}
C_{4}(y,z|x,x^{\prime }) &=&-\int_{u,v}C(z|x,u)\Gamma
_{3}^{tr}(y,u,v)G^{tr}(v,x^{\prime })-\int_{u,v}G^{tr}(x,u)\Gamma
_{3}^{tr}(y,u,v)C(z|v,x^{\prime })  \notag \\
&&-\int_{u,v}G^{tr}(x,u)\Gamma _{4}^{tr}(y,z,u,v)G^{tr}(v,x^{\prime }).
\label{C3}
\end{eqnarray}%
Here again the superscript ``tr" of $\Gamma _{3}^{tr}$ and $\Gamma _{4}^{tr}$
indicates they are derivatives of the truncated gap equation Eq.(\ref%
{gapeqoffshell}), i.e.,%
\begin{equation}
\Gamma _{3}^{tr}(y,u,v)=\left[ 6\varphi \left( u\right) \delta
(u-y)+3C(y|u,u)\right] \delta (u-v),  \label{C4}
\end{equation}%
and%
\begin{equation}
\Gamma _{4}^{tr}(y,z,u,v)=6\delta (u-z)\delta (u-y)\delta
(u-v)+3C_{4}(y,z|u,u)\delta (u-v)\text{.}  \label{C5}
\end{equation}%
Substituting Eq.(\ref{C4}) and Eq.(\ref{C5}) into Eq.(\ref{C3}), after some
rearrangements one finally gets
\begin{equation}
C_{4}(y,z|x,x^{\prime })+3\int_{u}G^{tr}(x,u)G^{tr}(u,x^{\prime
})C_{4}(y,z|u,u)=\cdots   \label{C6}
\end{equation}%
where ``$\cdots $" contains all the terms we have already encountered in
section 2. This chain equation for $C_{4}$ is analogous to Eq.(\ref{chaineqx}%
). We need only the particular case for $x=x^{\prime }.$ Using the
translation invariance, it can be easily solved in the momentum space:%
\begin{equation}
c_{p,q}^{4}=\frac{6\left[ 12\varphi ^{2}t_{p,q}-f_{q}\left( 1+3f_{-p}\right)
\left( 1+3f_{p+q}\right) \right] }{\left( 1+3f_{-p}\right) \left(
1+3f_{q}\right) \left( 1+3f_{p+q}\right) }\text{.}  \label{C7}
\end{equation}%
Here the Fourier transform is defined as%
\begin{equation}
C_{4}(y,z|x,x)=\int_{p,q}c_{p,q}^{4}e^{ip(y-z)}e^{iq(y-x)}\text{,}
\label{C8}
\end{equation}%
and $t_{p,q}$ is the ``triangle" integral%
\begin{equation}
t_{p,q}=\sum\nolimits_{k}g_{k}^{tr}g_{p-k}^{tr}g_{q+k}^{tr}.  \label{C9}
\end{equation}%
Substituting Eq.(\ref{C7}) and Eq.(\ref{chainsol}) into the Fourier form of
Eq.(\ref{C1}), one gets the final expression for $\Gamma _{3}$:%
\begin{equation}
\Gamma _{p,q}^{3}=6\varphi +3\left( c_{p+q}+c_{-q}+\varphi
c_{p,q}^{4}\right) .  \label{C10}
\end{equation}

The derivation of $\Gamma _{4}$ is similar but far more complicated\textit{.}
Except for the above mentioned chain functions, one would also need a higher
order chain,
\begin{equation}
C_{5}(y,z,u|x,x)\equiv \frac{\delta ^{2}G^{tr}\left( x,x\right) }{\delta
\varphi \left( y\right) \delta \varphi \left( z\right) \delta \varphi \left(
u\right) }=\int_{p,q,k}c_{p,q,k}^{5}e^{ip(y-z)}e^{iq(y-u)}e^{ik(y-x)},
\label{C11}
\end{equation}%
which is a complicated function of ``fish", ``triangle" and even ``box"
integrals,%
\begin{equation}
b_{p,q,k}=\sum%
\nolimits_{l}g_{l}^{tr}g_{p+l}^{tr}g_{p+q+l}^{tr}g_{p+q+k+l}^{tr}\text{.}
\label{C12}
\end{equation}%
The expression in terms of these integrals is rather bulky and will not be
presented here. The final expression for $\Gamma _{4}$ in terms of chain
functions is%
\begin{equation}
\Gamma _{p,q,k}^{4}=6+3\left(
c_{k,p+q}^{4}+c_{k,-p-k}^{4}+c_{p,q+k}^{4}+3\varphi c_{p,k,q}^{5}\right) .
\label{C13}
\end{equation}

Combining all these building blocks, one is able to calculate the
composite operator correlator
\begin{eqnarray}
P_{2}\left( k\right)  &\equiv &\int_{x}e^{-ikx}\left\langle \phi \left(
x\right) ^{2}\phi \left( 0\right) ^{2}\right\rangle =\sum_{pq}\left\langle
\phi _{p}\phi _{k-p}\phi _{q}\phi _{-k-q}\right\rangle   \label{P2composite}
\\
&=&C\delta _{k}+4\varphi ^{2}\left\langle \phi _{k}\phi _{-k}\right\rangle
_{c}+2\sum_{p}\left\langle \phi _{k-p}\phi _{p-k}\right\rangle
_{c}\left\langle \phi _{p}\phi _{-p}\right\rangle _{c}+4\varphi
\sum_{p}\left\langle \phi _{p}\phi _{k-p}\phi _{-k}\right\rangle
_{c}+\sum_{pq}\left\langle \phi _{p}\phi _{k-p}\phi _{q}\phi
_{-k-q}\right\rangle _{c}\text{.}  \notag
\end{eqnarray}%
The delta function part comes from disconnected diagrams and the three and
four - point connected correlation functions can be expressed via the two -
point connected correlator and the cumulants derived above. The results for $%
k\neq 0$ (that arise from the ``would be" broken phase solution of the
minimization equations) and that arise from the symmetric one are compared
in Figure 4 with the exact result of numerical diagonalization.

\section{Solution of minimization equations in the $O\left( 2\right) $ d=3
model}

\subsection{The ``fish" integral for the $O\left( 2\right) $ d=3 model}

The result of fish integral Eq.(\ref{fishO(2)}) in 3D is:
\begin{equation}
f_{k}^{ab}=\frac{1}{8\pi k}\left[ \arcsin \frac{k^{2}+m_{a}^{2}-m_{b}^{2}}{%
\sqrt{\left( k^{2}-m_{a}^{2}+m_{b}^{2}\right) ^{2}+4k^{2}m_{a}^{2}}}-\arcsin
\frac{-k^{2}+m_{a}^{2}-m_{b}^{2}}{\sqrt{\left(
k^{2}-m_{a}^{2}+m_{b}^{2}\right) ^{2}+4k^{2}m_{a}^{2}}}\right] \text{.}
\label{B1}
\end{equation}%
In particular, one gets in the limit of $k\rightarrow 0:$%
\begin{equation}
f_{0}^{ab}=\frac{1}{4\pi \left( m_{a}+m_{b}\right) }.  \label{B2}
\end{equation}

\subsection{Solution of minimization equations for the $O\left( 2\right) $
d=3 model}

The minimization equations for the broken phase $\varphi \neq 0$ are
\begin{eqnarray}
0 &=&a_{r}+\varphi ^{2}-\frac{3\omega }{4\pi }m_{1}-\frac{\omega }{4\pi }%
m_{2};  \label{B3.a} \\
m_{1}^{2} &=&a_{r}+3\varphi ^{2}-\frac{3\omega }{4\pi }m_{1}-\frac{\omega }{%
4\pi }m_{2};  \label{B3.b} \\
m_{2}^{2} &=&a_{r}+\varphi ^{2}-\frac{\omega }{4\pi }m_{1}-\frac{3\omega }{%
4\pi }m_{2}\text{.}  \label{B3.c}
\end{eqnarray}%
The first two equations give us%
\begin{equation}
m_{1}^{2}=2\varphi ^{2}=\left( \frac{2\pi }{\omega }m_{2}^{2}+m_{2}\right)
^{2}.  \label{B4}
\end{equation}%
Substituting this into Eq.(\ref{B3.c}) one has%
\begin{equation}
a_{r}=\frac{-2\pi ^{3}m_{2}^{4}-2\pi ^{2}\omega m_{2}^{3}+\pi \omega
^{2}m_{2}^{2}+\omega ^{3}m_{2}}{\pi \omega ^{2}}  \label{B5}
\end{equation}%
The right hand side of the above equation has a maximum of $0.03935\omega
^{2}$, above which there is no solution. Therefore the line $a_{r}\left(
\omega \right) =0.03935\omega ^{2}$ is the boundary of the symmetry broken
solution in the $a_{r}\thicksim \omega $ phase diagram. The quartic equation
Eq.(\ref{B5}) does not uniquely determine $m_{2}$ in terms of $a_{r}$. One
requires additional conditions, $m_{2}>0$ and $\partial a_{r}/\partial
m_{2}\leq 0$. Since for this branch the ``Higgs" excitation within the CGA is
\begin{equation}
m_{H}^{2}=m_{1}^{2}+\Delta \Gamma _{k=0}^{11}=\frac{2\pi \omega m_{1}^{2}}{%
8\pi ^{2}m_{1}m_{2}+3\pi \omega \left( m_{1}+m_{2}\right) +\omega }\frac{%
-\partial a_{r}}{\partial m_{2}},  \label{m2CGA}
\end{equation}%
the conditions are necessary for positivity of \ $m_{H}^{2}$. Consistently
the boundary of symmetry broken phase is specified by $\partial
a_{r}/\partial m_{2}=0.$

For the symmetric phase $\varphi =0,$ the minimization equations reduce to,%
\begin{equation}
m_{1}^{2}=m_{2}^{2}\equiv m^{2}=a_{r}-\frac{\omega }{\pi }m\text{,}
\label{B6}
\end{equation}%
giving rise to the solution
\begin{equation}
m=\frac{1}{2\pi }\left( \sqrt{\omega ^{2}+4\pi ^{2}a_{r}}-\omega \right)
\text{.}  \label{B7}
\end{equation}%
It exists for any $a_{r}>0$. Therefore in the small region $%
0<a_{r}<0.03935\omega ^{2}$ both the symmetric and asymmetric solutions
exist.

\section{Proof of Ward Identity for the current correlator}

The partition function,%
\begin{equation}
Z\equiv \int \mathcal{D}\Psi e^{-S}=\int \mathcal{D}\Psi \exp \left[ -\frac{1%
}{\omega }\int_{\mathbf{r}}\left\{ \frac{1}{2}\left\vert \nabla \Psi
\right\vert ^{2}+\frac{a}{2}\left\vert \Psi \right\vert ^{2}+\frac{1}{4}%
\left\vert \Psi \right\vert ^{4}\right\} \right] ,  \label{D1}
\end{equation}%
is invariant under the local unitary transformation%
\begin{equation}
\Psi (\mathbf{r})\rightarrow \Psi (\mathbf{r})e^{i\theta (\mathbf{r})}\text{.%
}  \label{D2}
\end{equation}%
The measure $\mathcal{D}\Psi $ is invariant under this transformation. This
gives%
\begin{equation}
Z=\int \mathcal{D}\Psi \exp \left[ -\frac{1}{\omega }\int_{\mathbf{r}}\left(
\mathbf{J}+\frac{1}{2}\left\vert \Psi \right\vert ^{2}\nabla \theta \right)
\cdot \nabla \theta \right] e^{-S}\text{.}  \label{D3}
\end{equation}%
For small $\nabla \theta $ the integrand in the above equation can be
expanded to the second order of $\nabla \theta $:%
\begin{equation}
1-\frac{1}{\omega }\int_{\mathbf{r}}\mathbf{J}\cdot \nabla \theta -\frac{1}{%
2\omega }\int_{\mathbf{r}}\left\vert \Psi \right\vert ^{2}\left( \nabla
\theta \right) ^{2}+\frac{1}{2\omega ^{2}}\int_{\mathbf{r,r}^{\prime
}}\left( \mathbf{J}\cdot \nabla \theta \right) _{\mathbf{r}}\left( \mathbf{J}%
\cdot \nabla \theta \right) _{\mathbf{r}^{\prime }}.  \label{D4}
\end{equation}%
Due to the invariance of $Z$, linear and quadratic terms in $\theta $
vanish:
\begin{eqnarray}
\nabla \cdot \left\langle \mathbf{J}\right\rangle  &=&0;  \label{D5} \\
\partial _{\mathbf{r}}^{\alpha }\partial _{\mathbf{r}^{\prime }}^{\beta
}\left( \frac{1}{\omega }\left\langle \mathbf{J}_{\mathbf{r}}^{\alpha }%
\mathbf{J}_{\mathbf{r}^{\prime }}^{\beta }\right\rangle -\left\langle
\left\vert \Psi _{\mathbf{r}}\right\vert ^{2}\right\rangle \delta _{\alpha
\beta }\delta \left( \mathbf{r-r}^{\prime }\right) \right)  &=&0\text{.}
\end{eqnarray}%
The Fourier transform of the second equation leads to the following ``Ward
Identity":%
\begin{equation}
0=k^{\alpha }k^{\beta }\left[ \frac{1}{\omega }\left\langle J_{\mathbf{k}%
}^{\alpha }J_{-\mathbf{k}}^{\beta }\right\rangle -\sum\nolimits_{\mathbf{q}%
}\left\langle \left\vert \Psi _{\mathbf{q}}\right\vert ^{2}\right\rangle
\delta _{\alpha \beta }\right] =k^{2}\left[ \frac{1}{\omega }\chi
_{L}(k^{2})-\sum\nolimits_{\mathbf{q}}\left\langle \left\vert \Psi _{\mathbf{%
q}}\right\vert ^{2}\right\rangle \right] \text{.}
\end{equation}

\end{document}